\documentclass[letterpaper,american,prl,simglecolumn,amsmath,amssymb,showpacs,superscriptaddress]{revtex4-1}
\usepackage[T1]{fontenc}
\usepackage{babel}
\usepackage{prettyref}
\usepackage{amstext}
\usepackage{graphicx}
\usepackage[unicode=true,
 bookmarks=true,bookmarksnumbered=false,bookmarksopen=false,
 breaklinks=false,pdfborder={0 0 1},backref=false,colorlinks=false]
 {hyperref}
\hypersetup{pdftitle={Functional methods for disordered neural networks},
 pdfauthor={Schuecker, Goedeke, Dahmen, Helias},
 colorlinks=true,linkcolor=black,citecolor=black,urlcolor=black,filecolor=black}
\usepackage{breakurl}

\makeatletter

\newcommand{\lyxdot}{.}

\usepackage{MnSymbol}  
\usepackage{simplewick}  

\newrefformat{cap}{\hyperref[#1]{Figure~\ref{#1}}}
\newrefformat{fig}{\hyperref[#1]{Figure~\ref{#1}}}
\newrefformat{tab}{\hyperref[#1]{Table ~\ref{#1}}}
\newrefformat{sec}{\hyperref[#1]{Section~\ref{#1}}}
\newrefformat{sub}{\hyperref[#1]{Section~\ref{#1}}}
\newrefformat{cha}{\hyperref[#1]{Chapter~\ref{#1}}}

\makeatother

\begin{document}
\global\long\def\D{\mathcal{D}}
\global\long\def\bx{\mathbf{x}}
\global\long\def\bl{\mathbf{l}}
\global\long\def\bh{\mathbf{h}}
\global\long\def\bJ{\mathbf{J}}
\global\long\def\N{\mathcal{N}}
\global\long\def\hh{\hat{h}}
\global\long\def\bhh{\mathbf{\hh}}
\global\long\def\T{\mathrm{T}}
\global\long\def\by{\mathrm{\mathbf{y}}}
\global\long\def\diag{\mathrm{diag}}
\global\long\def\Ftr#1#2{\mathfrak{F}\left[#1\right]\left(#2\right)}
\global\long\def\iFtr#1#2{\mathfrak{F^{-1}}\left[#1\right]\left(#2\right)}
\global\long\def\D{\mathcal{D}}
\global\long\def\T{\mathrm{T}}
\global\long\def\Gammafl{\Gamma_{\mathrm{fl}}}
\global\long\def\gammafl{\gamma_{\mathrm{fl}}}
\global\long\def\E#1{\left\langle #1\right\rangle }

\global\long\def\D{\mathcal{D}}
\global\long\def\J{\mathbf{J}}
\global\long\def\one{\mathbf{1}}
\global\long\def\e{\mathbf{e}}
\global\long\def\Cpp{\mathcal{K}_{\phi^{\prime}\phi^{\prime}}^{(0)}}
\global\long\def\CCpp{C_{\phi^{\prime}\phi^{\prime}}^{(0)}}
\global\long\def\Cppj{\mathcal{K}_{\phi_{j}^{\prime}\phi_{j}^{\prime}}^{(0)}}
\global\long\def\tx{\tilde{x}}
\global\long\def\tj{\tilde{j}}
\global\long\def\xo{x^{(0)}}
\global\long\def\xii{x^{(1)}}
\global\long\def\txi{\tilde{x}^{(1)}}
\global\long\def\bx{\mathbf{x}}
\global\long\def\tbx{\tilde{\mathbf{x}}}
\global\long\def\bl{\mathbf{j}}
\global\long\def\tbj{\tilde{\mathbf{j}}}
\global\long\def\bk{\mathbf{k}}
\global\long\def\tbk{\tilde{\mathbf{k}}}
\global\long\def\bh{\mathbf{h}}
\global\long\def\bJ{\mathbf{J}}
\global\long\def\bN{\mathcal{N}}
\global\long\def\bH{\mathbf{H}}
\global\long\def\bK{\mathbf{K}}
\global\long\def\bxo{\bx^{(0)}}
\global\long\def\tbxo{\tilde{\bx}^{(0)}}
\global\long\def\bxi{\bx^{(1)}}
\global\long\def\tbxi{\tilde{\bx}^{(1)}}
\global\long\def\tbxi{\tbx^{(1)}}
\global\long\def\tpsi{\tilde{\psi}}
\global\long\def\Cxi{C_{x^{(1)}x^{(1)}}}
\global\long\def\bxxi{\mathbf{\xi}}
\global\long\def\N{\mathcal{N}}
\global\long\def\bW{\mathbf{W}}
\global\long\def\bon{\mathbf{1}}

\global\long\def\mcC{\mathcal{C}}

\title{Functional methods for disordered neural networks}

\author{Jannis Sch\"ucker}

\affiliation{Institute of Neuroscience and Medicine (INM-6) and Institute for
Advanced Simulation (IAS-6) and JARA BRAIN Institute I, Jülich Research
Centre, Jülich, Germany}

\author{Sven Goedeke}

\affiliation{Institute of Neuroscience and Medicine (INM-6) and Institute for
Advanced Simulation (IAS-6) and JARA BRAIN Institute I, Jülich Research
Centre, Jülich, Germany}

\author{David Dahmen}

\affiliation{Institute of Neuroscience and Medicine (INM-6) and Institute for
Advanced Simulation (IAS-6) and JARA BRAIN Institute I, Jülich Research
Centre, Jülich, Germany}

\author{Moritz Helias}

\affiliation{Institute of Neuroscience and Medicine (INM-6) and Institute for
Advanced Simulation (IAS-6) and JARA BRAIN Institute I, Jülich Research
Centre, Jülich, Germany}

\affiliation{Department of Physics, Faculty 1, RWTH Aachen University, Aachen,
Germany}
\email{m.helias@fz-juelich.de}

\selectlanguage{american}%

\date{\today}

\pacs{87.19.lj, 64.60.an, 75.10.Nr, 05.40.-a}
\begin{abstract}
Neural networks of the brain form one of the most complex systems
we know. Many qualitative features of the emerging collective phenomena,
such as correlated activity, stability, response to inputs, chaotic
and regular behavior, can, however, be understood in simple models
that are accessible to a treatment in statistical mechanics, or, more
precisely, classical statistical field theory.

This tutorial presents the fundamentals behind contemporary developments
in the theory of neural networks of rate units \citep[e.g. ][]{Aljadeff15_088101,Stern14_062710,Hermann12_018702}
that are based on methods from statistical mechanics of classical
systems with a large number of interacting degrees of freedom. In
particular we will focus on a relevant class of systems that have
quenched (time independent) disorder. In neural networks, the main
source of disorder arises from random synaptic couplings between neurons.
These systems are in many respects similar to spin glasses \citep{Fischer91}.
The tutorial therefore also explains the methods for these disordered
systems as far as they are applied in neuroscience.

The presentation consists of two parts. In the first part we introduce
stochastic differential equations (in the Ito-formulation and Stratonovich
formulation) and present their treatment in the Martin\textendash Siggia\textendash Rose-De
Dominicis path integral formalism \citep{Martin73,DeDomincis78_353},
reviewed in \citep{Altland01,Chow15,hertz2016_arxiv}. In the second
part we will employ this language to derive the dynamic mean-field
theory for deterministic random networks \citep{Sompolinsky88_259}.
To our knowledge, a detailed presentation of the methods behind the
results of this seminal paper is still lacking in the literature.
Any inaccuracies in the present manuscript should therefore not be
attributed to the authors of the original work \citep{Sompolinsky88_259},
but to those of this tutorial. In deriving the formalism, we will
follow the De Dominicis approach \citep{DeDomincis78_353}, that was
also employed to obtain the dynamic mean-field theory of spin glasses
\citep{Sompolinsky81,Sompolinsky82_6860,Crisanti87_4922}. The formalism
in particular explains the statistics of the fluctuations in these
networks and the emergence of different phases with regular and chaotic
dynamics \citep{Sompolinsky88_259}. We will also cover a recent extension
of the model to stochastic units \citep{Goedeke16_arxiv}.

\vfill{}
\pagebreak{}
\end{abstract}
\maketitle
\tableofcontents{}

\pagebreak{}

\section{Functional formulation of stochastic differential equations\label{sec:Martin-Siggia-Rose-De-Dominicis}}

We here follow \citet{Chow10_1009} to derive the Martin-Siggia-Rose
\citep{Martin73,janssen1976_377,dedominicis1976_247,DeDomincis78_353,Altland01,Chow15}
path integral representation of a stochastic differential equation
and \citet{Wio89_7312} to obtain the Onsager-Machlup path integral.
We generalize the notation to also include the Stratonovich convention
as in \citep{Wio89_7312}. \citet{hertz2016_arxiv} also provide a
pedagogical survey of the Martin-Siggia-Rose path integral formalism
for the dynamics of stochastic and disordered systems.

The presented functional formulation of dynamics is advantageous in
several respects. First, it recasts the dynamical equations into a
path-integral, where the dynamic equations give rise to the definition
of an ``action''. In this way, the known tools from theoretical
physics, such as perturbation expansions with the help of Feynman
diagrams or the loopwise expansions to obtain a systematic treatment
of fluctuations \citep{ZinnJustin96}, can be applied. Within neuroscience,
the recent review \citep{Chow15} illustrates the first, the work
by \citep{Buice07_051919} the latter approach. Moreover, this formulation
will be essential for the treatment of disordered systems in \prettyref{sec:Sompolinsky-Crisanti-Sommers-theory},
following the spirit of the work by \citet{DeDomincis78_353} to obtain
a generating functional that describes an average system belonging
to an ensemble of systems with random parameters.

Many dynamic phenomena can be described by differential equations.
Often, the presence of fluctuations is represented by an additional
stochastic forcing. We therefore consider the \textbf{stochastic differential
equation} (SDE) 
\begin{eqnarray}
dx(t) & = & f(x)\,dt+dW(t)\label{eq:SDE}\\
x(0+) & = & a,\nonumber 
\end{eqnarray}
where $a$ is the initial value and $dW$ a stochastic increment.
Stochastic differential equations are defined as the limit $h\to0$
of a dynamics on a discrete time lattice of spacing $h$. For discrete
time $t_{l}=lh$, $l=0,\ldots,M$, the solution of the SDE consists
of the discrete set of points $x_{l}=x(t_{l})$. For the discretization
there are mainly two conventions used, the Ito and the Stratonovich
convention \citep{Gardiner09}. Since we only consider additive noise,
i.e. the stochastic increment in \eqref{eq:SDE} does not depend on
the state, both conventions yield the same continuous-time limit.
However, as we will see, different discretization conventions of the
drift term lead to different path integral representations. The Ito
convention defines the symbolic notation of \eqref{eq:SDE} to be
interpreted as 
\begin{eqnarray*}
x_{i+1}-x_{i} & = & f(x_{i})\,h+a\delta_{i0}+W_{i},
\end{eqnarray*}
where $W_{i}$ is a stochastic increment that follows a probabilistic
law. A common choice for $W_{i}$ is a normal distribution $\rho(W_{i})=\N(0,\,hD)$,
called a Wiener increment. Here the parameter $D$ controls the variance
of the noise. The term $a\delta_{i0}$ ensures that the solution obeys
the stated initial condition, assuming that $x_{i\le0}=0$ in the
absence of noise $W_{0}=0$. If the variance of the increment is proportional
to the time step $h$, this amounts to a $\delta$-distribution in
the autocorrelation of the noise $\xi=\frac{dW}{dt}$. The Stratonovich
convention, also called mid-point rule, instead interprets the SDE
as
\begin{eqnarray*}
x_{i+1}-x_{i} & = & f\left(\frac{x_{i+1}+x_{i}}{2}\right)\,h+a\delta_{i0}+W_{i}.
\end{eqnarray*}
Both conventions can be treated simultaneously by defining
\begin{eqnarray}
x_{i+1}-x_{i} & = & f(\mbox{\ensuremath{\alpha}}x_{i+1}+(1-\alpha)x_{i})\,h+a\delta_{i0}+W_{i}\label{eq:discrete_sde}\\
\alpha & \in & [0,1].\nonumber 
\end{eqnarray}
Here $\alpha=0$ corresponds to the Ito convention and $\alpha=\frac{1}{2}$
to Stratonovich. If the noise is drawn independently for each time
step, i.e. if it is white, the probability density of the path $x(t)$,
i.e. a distribution in the points $x_{1},\ldots,x_{M}$, can be written
as

\begin{eqnarray}
p(x_{1},\ldots,x_{M}|a) & \equiv & \int\Pi_{i=0}^{M-1}dW_{i}\,\rho(W_{i})\,\delta(x_{i+1}-y_{i+1}(W_{i},x_{i})),\label{eq:p_of_x}
\end{eqnarray}
where, by \eqref{eq:discrete_sde}, $y_{i+1}(W_{i},x_{i})$ is understood
as the solution of \eqref{eq:discrete_sde} at time point $i+1$ given
the noise realization $W_{i}$ and the solution until the previous
time point $x_{i}$: The solution of the SDE starts at $i=0$ with
$x_{0}=0$ so that $W_{0}$ and $a$ together determine $x_{1}$.
In the next time step, $W_{1}$ and $x_{1}$ together determine $x_{2}$,
and so on. In the Ito-convention ($\alpha=0$) we have an explicit
solution $y_{i+1}(W_{i},x_{i})=x_{i}+f(x_{i})\,h+a\delta_{i0}+W_{i}$,
while the Stratonovich convention yields an implicit equation, since
$x_{i+1}$ appears as an argument of $f$. We will see in in \eqref{eq:normalization_strato}
that the latter gives rise to a non-trivial normalization factor $(1-\alpha f^{\prime}h)$
for $p$, while for the former this factor is unity.

The notation $y_{i+1}(W_{i},x_{i})$ indicates that the solution only
depends on the last time point $x_{i}$, but not on the history longer
ago, which is called the \textbf{Markov property} of the process.
This form also shows that the density is correctly normalized, because
integrating over all paths
\begin{align*}
 & \int dx_{1}\,\cdots\int dx_{M}\,p(x_{1},\ldots,x_{M}|a)=\int\Pi_{i=0}^{M-1}dW_{i}\rho(W_{i})\,\underbrace{\int dx_{i+1}\,\delta(x_{i+1}-y_{i+1}(W_{i},x_{i}))}_{=1}\\
= & \Pi_{i=0}^{M-1}\int dW_{i}\rho(W_{i})=1
\end{align*}
yields the normalization condition of $\rho(W_{i})$, $i=0,\ldots,M-1$,
the distribution of the stochastic increments. In the limit $M\to\infty$,
we therefore define the probability functional as $p[x|a]:=\lim_{M\to\infty}p(x_{1},\ldots,x_{M}|a)$.

Using \eqref{eq:p_of_x} and the substitution $\delta(y)\,dy=\delta(\phi(x_{i+1}))\phi^{\prime}dx_{i+1}$
with $y=\phi(x_{i+1})=W_{i}(x_{i+1})$ obtained by solving \eqref{eq:discrete_sde}
for $W_{i}$

\begin{eqnarray}
W_{i}(x_{i+1}) & = & x_{i+1}-x_{i}-f(\mbox{\ensuremath{\alpha}}x_{i+1}+(1-\alpha)x_{i})\,h-a\delta_{i0}\nonumber \\
\frac{\partial W_{i}}{\partial x_{i+1}} & =\phi^{\prime}= & 1-\alpha f^{\prime}h\label{eq:normalization_strato}
\end{eqnarray}
we obtain

\begin{eqnarray}
p(x_{1},\ldots,x_{M}|a) & = & \int\Pi_{i=0}^{M-1}dW_{i}\,\rho(W_{i})\,\times\label{eq:density}\\
 & \times & \delta(W_{i}-\underbrace{x_{i+1}-x_{i}-f(\mbox{\ensuremath{\alpha}}x_{i+1}+(1-\alpha)x_{i})\,h-a\delta_{i0}}_{W_{i}(x_{i+1})})\,(1-\alpha f^{\prime}h).\nonumber \\
 & = & \Pi_{i=0}^{M-1}\rho(x_{i+1}-x_{i}-f(\mbox{\ensuremath{\alpha}}x_{i+1}+(1-\alpha)x_{i})\,h-a\delta_{i0})\,\left(1-\alpha h\,f^{\prime}(\mbox{\ensuremath{\alpha}}x_{i+1}+(1-\alpha)x_{i})\right).\nonumber 
\end{eqnarray}

In section \prettyref{sub:Onsager-Machlup-path-integral} we will
look at the special case of Gaussian noise and derive the so called
Onsager-Machlup path integral \citep{Onsager53}. This path integral
has a square in the action, originating from the Gaussian noise. For
many applications, this square complicates the analysis of the system.
The formulation presented in \prettyref{sub:Martin-Siggia-Rose-De-Dominicis}
removes this square on the expense of the introduction of an additional
field, the so called response field. This formulation has the additional
advantage that responses of the system to perturbations can be calculated
in compact form, as we will see below.

\subsection{Onsager-Machlup path integral\label{sub:Onsager-Machlup-path-integral}}

For the case of a Gaussian noise $\rho(W_{i})=\N(0,\,Dh)=\frac{1}{\sqrt{2\pi Dh}}\,e^{-\frac{W_{i}^{2}}{2Dh}}$
the variance of the increment is 
\begin{eqnarray}
\langle W_{i}W_{j}\rangle & = & \begin{cases}
Dh & \quad i=j\\
0 & \quad i\neq j
\end{cases}\label{eq:GWN_OM}\\
 & = & \delta_{ij}\,Dh.\nonumber 
\end{eqnarray}
Using the Gaussian noise and then taking the limit $M\to\infty$ of
eq. \eqref{eq:density} with $1-\alpha f^{\prime}h\to\exp(-\alpha f^{\prime}h)$
we obtain
\begin{eqnarray*}
p(x_{1},\ldots,x_{M}|a) & = & \Pi_{i=0}^{M-1}\rho(x_{i+1}-x_{i}-f(\mbox{\ensuremath{\alpha}}x_{i+1}+(1-\alpha)x_{i})\,h-a\delta_{i0})\,(1-\alpha f^{\prime}h)+O(h^{2})\\
 & = & \Pi_{i=0}^{M-1}\frac{1}{\sqrt{2\pi Dh}}\,\exp\left[-\frac{1}{2Dh}(x_{i+1}-x_{i}-f(\mbox{\ensuremath{\alpha}}x_{i+1}+(1-\alpha)x_{i})\,h-a\delta_{i0})^{2}-\alpha f^{\prime}h\right]+O(h^{2})\\
 & = & \left(\frac{1}{\sqrt{2\pi Dh}}\right)^{M}\,\exp\left[-\frac{1}{2D}\sum_{i=0}^{M-1}\left[(\frac{x_{i+1}-x_{i}}{h}-f(\mbox{\ensuremath{\alpha}}x_{i+1}+(1-\alpha)x_{i})-a\frac{\delta_{i0}}{h})^{2}-\alpha f^{\prime})\right]h\right]+O(h^{2}).
\end{eqnarray*}
We will now define a symbolic notation by recognizing $\lim_{h\to0}\frac{x_{i+1}-x_{i}}{h}=\partial_{t}x(t)$
as well as $\lim_{h\to0}\frac{\delta_{i0}}{h}=\delta(t)$ and $\lim_{h\to0}\sum_{i}f(hi)\,h=\int f(t)\,dt$
\begin{align}
p[x|x(0+)=a]\,\mathcal{D}_{\sqrt{2\pi Dh}}x & =\exp\left(-\frac{1}{2D}\int_{0}^{T}(\partial_{t}x-f(x)-a\delta(t))^{2}-\alpha f^{\prime}\,dt\right)\D_{\sqrt{2\pi Dh}}x\label{eq:OM_pathint}\\
 & :=\lim_{M\to\infty}p(x_{1},\ldots,x_{M}|a)\frac{dx_{1}}{\sqrt{2\pi Dh}}\ldots\frac{dx_{M}}{\sqrt{2\pi Dh}},\nonumber 
\end{align}
where we defined the integral measure $\mathcal{D}_{\sqrt{2\pi Dh}}x:=\Pi_{i=1}^{M}\frac{dx_{i}}{\sqrt{2\pi Dh}}$
to obtain a normalized density $1=\int\mathcal{D}_{\sqrt{2\pi Dh}}x\,p[x|x(0+)=a]$.

\subsection{Martin-Siggia-Rose-De Dominicis-Janssen (MSRDJ) path integral\label{sub:Martin-Siggia-Rose-De-Dominicis}}

The square in the action \prettyref{eq:OM_pathint} sometimes has
disadvantages for analytical reasons, for example if quenched averages
are to be calculated, as we will do in \prettyref{sec:Sompolinsky-Crisanti-Sommers-theory}.
To avoid the square we will here introduce an auxiliary field, the
\textbf{response field} $\tx$ (the name will become clear in \prettyref{sub:Response-function-in}).
This field enters the probability functional \prettyref{eq:density}
by representing the $\delta$-distribution by its Fourier integral
\begin{eqnarray}
\delta(x) & = & \frac{1}{2\pi i}\int_{-i\infty}^{i\infty}\,d\tilde{x}\,e^{\tilde{x}x}.\label{eq:Fourier_delta}
\end{eqnarray}
Replacing the $\delta$-distribution at each time slice by an integral
over $\tx_{i}$ at the corresponding slice, eq. \eqref{eq:density}
takes the form 
\begin{eqnarray}
p(x_{1},\ldots,x_{M}|a) & = & \Pi_{i=0}^{M-1}\left\{ \int dW_{i}\rho(W_{i})\,\int_{-i\infty}^{i\infty}\frac{d\tilde{x}_{i}}{2\pi i}\,\exp\left(\tilde{x}_{i}(x_{i+1}-x_{i}-f(\alpha x_{i+1}+(1-\alpha)x_{i})h-W_{i}-a\delta_{i0})-\alpha f^{\prime}h\right)\right\} \nonumber \\
 & = & \Pi_{i=0}^{M-1}\left\{ \int_{-i\infty}^{i\infty}\frac{d\tilde{x}_{i}}{2\pi i}\,\exp\left(\tilde{x}_{i}(x_{i+1}-x_{i}-f(\alpha x_{i+1}+(1-\alpha)x_{i})h-a\delta_{i0})-\alpha f^{\prime}h\right)\,Z_{W}(-\tilde{x}_{i})\right\} \label{eq:general_path_prob}\\
\nonumber \\
Z_{W}(-\tilde{x}) & \equiv & \int dW_{i}\rho(W_{i})\,e^{-\tilde{x}W_{i}}=\langle e^{-\tilde{x}W_{i}}\rangle_{W_{i}}.\nonumber 
\end{eqnarray}
Here $Z_{W}(-\tilde{x})$ is the moment generating function \citep{Gardiner85}
also known as the characteristic function of the noise process, which
is identical to the Fourier transform of the density (with $i\omega=-\tilde{x}$).
Note the index $i$ of the field $\tilde{x}_{i}$ is the same as the
index of the noise variable $W_{i}$, which allows the definition
of the characteristic function $Z_{W}$. Hence the distribution of
the noise only appears in the probability functional in the form of
$Z_{W}(-\tx)$. For Gaussian noise \prettyref{eq:GWN_OM} the characteristic
function is
\begin{eqnarray}
Z_{W}(-\tilde{x}) & = & \frac{1}{\sqrt{2\pi Dh}}\int\,dW\,e^{-\frac{W^{2}}{2Dh}}\,e^{-\tilde{x}W}=\frac{1}{\sqrt{2\pi Dh}}\int\,dW\,e^{-\frac{1}{2Dh}(W+Dh\tilde{x})^{2}}\,e^{\frac{Dh}{2}\tilde{x}^{2}}\nonumber \\
 & = & e^{\frac{Dh}{2}\tilde{x}^{2}}.\label{eq:Gaussian_generator_discrete}
\end{eqnarray}

\subsection{Moment generating functional}

The probability distribution \eqref{eq:general_path_prob} is a distribution
for the random variables $x_{1},\ldots,x_{M}$. We can alternatively
describe the probability distribution by the moment-generating functional
by adding the terms $\sum_{l=1}^{M}j_{l}x_{l}h$ to the action and
integrating over all paths
\begin{eqnarray}
Z(j_{1},\ldots,j_{M}) & := & \Pi_{l=1}^{M}\left\{ \int_{-\infty}^{\infty}dx_{l}\,\exp\left(j_{l}x_{l}h\right)\right\} \,p(x_{1},\ldots,x_{M}|a).\label{eq:generating_functional_def}
\end{eqnarray}
Moments of the path can be obtained by taking derivatives (writing
$\bl=(j_{1},\ldots,j_{M})$)

\begin{eqnarray}
\left.\frac{\partial}{\partial(h\,j_{k})}Z(\bl)\right|_{\bl=0} & = & \Pi_{l=1}^{M}\left\{ \int_{-\infty}^{\infty}dx_{l}\right\} \,p(x_{1},\ldots,x_{M}|a)\,x_{k}\nonumber \\
 & \equiv & \langle x_{k}\rangle.\label{eq:moments_func_deriv}
\end{eqnarray}
For $M\to\infty$ and $h\to0$ the additional term $\exp\left(\sum_{l=1}^{M}j_{l}\,x_{l}h\right)\stackrel{h\to0}{\rightarrow}\exp\left(\int j(t)x(t)\,dt\right)$.
So the derivative on the left hand side of \prettyref{eq:moments_func_deriv}
turns into the functional derivative 
\begin{align*}
\frac{\partial}{\partial(hj_{k})}Z(\bl) & \equiv\lim_{\epsilon\to0}\frac{1}{\epsilon}\left(Z(j_{1},\ldots,j_{k}+\frac{\epsilon}{h},\,j_{k+1},\ldots,j_{M}]-Z(j_{1},\ldots,j_{k},\ldots,j_{M})\right)\stackrel{h\to0}{\rightarrow}\frac{\delta}{\delta j(t)}Z[j],
\end{align*}
and the moment becomes $\langle x(t)\rangle$ at time point $t=hk$.
The generating functional takes the explicit form
\begin{eqnarray}
Z(\bl) & = & \Pi_{l=1}^{M}\left\{ \int_{-\infty}^{\infty}dx_{l}\exp\left(j_{l}x_{l}h\right)\right\} \Pi_{k=0}^{M-1}\left\{ \int_{-i\infty}^{i\infty}\frac{d\tilde{x}_{k}}{2\pi i}\,Z_{W}(-\tilde{x}_{k})\right\} \,\times\label{eq:generating_functional_discrete}\\
 &  & \times\exp\left(\sum_{l=0}^{M-1}\tilde{x}_{l}(x_{l+1}-x_{l}-f(\alpha x_{l+1}+(1-\alpha)x_{l})h-a\delta_{l0})-\alpha f^{\prime}h\right).\nonumber 
\end{eqnarray}
Note the different index ranges for the path coordinates $x_{1},\ldots,x_{M}$
and the response field $\tilde{x}_{0},\ldots,\tilde{x}_{M-1}$. Letting
$h\to0$ we now define the path integral as the generating functional
\eqref{eq:generating_functional_discrete} and introduce the notations
$\Pi_{i=1}^{M}\int_{-\infty}^{\infty}dx_{i}\stackrel{h\to0}{\to}\int\mathcal{D}x$
as well as $\Pi_{i=0}^{M-1}\int_{-i\infty}^{i\infty}\frac{d\tilde{x}_{i}}{2\pi i}\stackrel{h\to0}{\to}\int\mathcal{D}_{2\pi i}\tilde{x}$.
Note that the different index ranges and the different integral boundaries
are implicit in this notation, depending on whether we integrate over
$x(t)$ or $\tilde{x}(t)$. We hence write symbolically for the probability
distribution \eqref{eq:general_path_prob}
\begin{eqnarray}
p[x(t)|x(0+)=a] & = & \int\mathcal{D}_{2\pi i}\tilde{x}\,\exp\left(\int_{-\infty}^{\infty}\tilde{x}(t)(\partial_{t}x-f(x)-a\delta(t))-\alpha f^{\prime}\,dt\right)\,Z_{W}[-\tilde{x}]\label{eq:martin_siggia_rose_general}\\
 & = & \int\mathcal{D}_{2\pi i}\tilde{x}\,\exp\left(\tilde{x}^{\T}(\partial_{t}x-f(x)-a\delta(t))-\int_{-\infty}^{\infty}\alpha f^{\prime}\,dt\right)\,Z_{W}[-\tilde{x}]\nonumber \\
\nonumber \\
Z_{W}[-\tilde{x}_{k}] & = & \left\langle \exp\left(-\int_{-\infty}^{\infty}\tilde{x}(t)\,dW(t)\right)\right\rangle {}_{W}\nonumber \\
 & = & \left\langle \exp\left(-\tilde{x}^{\T}dW\right)\right\rangle {}_{W}\nonumber 
\end{eqnarray}
where the respective second lines use the definition of the inner
product on the space of functions 
\begin{align}
x^{\T}y & :=\int_{-\infty}^{\infty}x(t)y(t)\,dt.\label{eq:inner_product}
\end{align}
This vectorial notation also reminds us of the discrete origin of
the path integral. Note that the lattice derivative appearing in \prettyref{eq:martin_siggia_rose_general}
follows the definition $\partial_{t}x=\lim_{h\to0}\frac{1}{h}\left(x_{t/h+1}-x_{t/h}\right)$.
We compactly denote the generating functional \eqref{eq:generating_functional_discrete}
as

\begin{align}
Z[j] & =\int\D x\,\int\D_{2\pi i}\tilde{x}\,\exp\left(\int\tilde{x}(t)(\partial_{t}x-f(x)-a\delta(t))-\alpha f^{\prime}+j(t)x(t)\,dt\right)\,\,Z_{W}[-\tilde{x}].\label{eq:MSR_Z_tilde_J}
\end{align}
For Gaussian white noise we have with \eqref{eq:Gaussian_generator_discrete}
the moment generating functional $\,Z_{W}[-\tilde{x}]=\exp\left(\frac{D}{2}\,\tilde{x}^{\T}\tx\right)$.
If in addition, we adopt the Ito convention, i.e. setting $\alpha=0$,
we get
\begin{align}
Z[j] & =\int\D x\,\int\D_{2\pi i}\tilde{x}\,\exp\left(\tilde{x}^{\T}(\partial_{t}x-f(x)-a\delta(t))+\frac{D}{2}\tx^{\T}\tx+j^{\T}x\right).\label{eq:msr_gwn}
\end{align}

\subsection{Response function in the MSRDJ formalism\label{sub:Response-function-in}}

The path integral \eqref{eq:general_path_prob} can be used to determine
the response of the system to an external perturbation. To this end
we consider the stochastic differential equation \eqref{eq:SDE} that
is perturbed by a time-dependent drive $-\tilde{j}(t)$

\begin{eqnarray*}
dx(t) & = & (f(x)-\tilde{j}(t))\,dt+dW(t)\\
x(0+) & = & a.
\end{eqnarray*}

In the following we will only consider the Ito convention and set
$\alpha=0$. We perform the analogous calculation that leads from
\eqref{eq:SDE} to \eqref{eq:generating_functional_discrete} with
the additional term $-\tilde{j}(t)$ due to the perturbation. In the
sequel we will see that, instead of treating the perturbation explicitly,
it can be expressed with the help of a second source term. The generating
functional including the perturbation is

\begin{align}
Z(\bl,\tbj) & =\Pi_{l=1}^{M}\left\{ \int_{-\infty}^{\infty}dx_{l}\right\} \Pi_{k=0}^{M-1}\left\{ \int_{-i\infty}^{i\infty}\frac{d\tilde{x}_{k}}{2\pi i}\,Z_{W}(-\tilde{x}_{k})\right\} \,\times\nonumber \\
 & \times\exp\left(\sum_{l=0}^{M-1}\tilde{x}_{l}(x_{l+1}-x_{l}-f(x_{l})h-a\delta_{l,0})+j_{l+1}x_{l+1}h+\tilde{x}_{l}\tilde{j}_{l}h\right)\label{eq:msr_z}\\
 & =\int\D x\,\int\D_{2\pi i}\tilde{x}\,Z_{W}[-\tilde{x}]\,\exp\left(\int_{-\infty}^{\infty}\tilde{x}(t)(\partial_{t}x-f(x)-a\delta(t))+j(t)x(t)+\tilde{j}(t)\tilde{x}(t)\,dt\right),\nonumber 
\end{align}
where we moved the $\tilde{j}-$dependent term out of the parenthesis.

Note that the external field $j$ is indexed from $1,\ldots,M$ (as
$x_{l}$) whereas $\tilde{j}$ is indexed $0,\ldots,M-1$ (as $\tilde{x}$).
As before, the moments of the process follow as functional derivatives
\eqref{eq:moments_func_deriv} $\left.\frac{\delta}{\delta j(t)}Z[j,\tilde{j}]\right|_{j=\tilde{j}=0}=\langle x(t)\rangle$.
Higher order moments follow as higher derivatives.

The additional dependence on $\tilde{j}$ allows us to investigate
the response of arbitrary moments to a small perturbation localized
in time, i.e. $\tilde{j}(t)=-\epsilon\delta(t-s)$. In particular,
we characterize the average response of the first moment with respect
to the unperturbed system by the \textbf{response function} $\chi(t,s)$

\begin{eqnarray}
\chi(t,s) & := & \lim_{\epsilon\to0}\frac{1}{\epsilon}\left(\langle x(t)\rangle_{\tilde{j}=-\epsilon\delta(\cdot-s)}-\langle x(t)\rangle_{\tilde{j}=0}\right)\label{eq:MSR_response}\\
 & = & \lim_{\epsilon\to0}\frac{1}{\epsilon}\int\D x\,x(t)\,(p[x|\tilde{j}=-\epsilon\delta(t-s)]-p[x|\tilde{j}=0])\nonumber \\
 & = & \lim_{\epsilon\to0}\frac{1}{\epsilon}\left.\frac{\delta}{\delta j(t)}\left(Z[j,\tilde{j}-\epsilon\delta(t-s)]-Z[j,\tilde{j}]\right)\right|_{j=\tilde{j}=0}\nonumber \\
 & = & \left.-\frac{\delta}{\delta j(t)}\frac{\delta}{\delta\tilde{j}(s)}Z[j,\tilde{j}]\right|_{j=\tilde{j}=0}\nonumber \\
 & = & -\langle x(t)\,\tilde{x}(s)\rangle,\nonumber 
\end{eqnarray}
where we used the definition of the functional derivative from the
third to the fourth line. So instead of treating a small perturbation
explicitly, the response of the system to a perturbation can be obtained
by a functional derivative with respect to $\tilde{j}$: $\tilde{j}$
couples to $\tilde{x}$, $\tilde{j}$ contains perturbations, therefore
$\tilde{x}$ measures the response and is the so called response field.
The response function $\chi(t,s)$ can then be used as a kernel to
obtain the mean response of the system to a small external perturbation
of arbitrary temporal shape.

There is an important difference for the response function between
the Ito and Stratonovich formulation, that is exposed in the time-discrete
formulation. For the perturbation $\tilde{j}(t)=-\epsilon\delta(t-s)$,
we obtain the perturbed equation, where $\frac{s}{h}$ denotes the
discretized time point at which the perturbation is applied. The perturbing
term must be treated analogously to $f$, so 
\begin{eqnarray*}
x_{i+1}-x_{i} & = & f(\mbox{\ensuremath{\alpha}}x_{i+1}+(1-\alpha)x_{i})\,h+\epsilon\left(\alpha\delta_{i+1,\frac{s}{h}}+(1-\alpha)\delta_{i,\frac{s}{h}}\right)+W_{i}\\
\alpha & \in & [0,1].
\end{eqnarray*}
Consequently, the value of the response function $\chi(s,s)$ at the
time of the perturbation depends on the choice of $\alpha$. We denote
as $x_{j}^{\epsilon}$ the solution after application of the perturbation,
as $x_{j}^{0}$ the solution without; for $i<j$ the two are identical
and the equal-time response is
\begin{eqnarray*}
\chi(s,s) & = & \lim_{\epsilon\to0}\frac{1}{\epsilon}\left(x_{\frac{s}{h}}^{\epsilon}-x_{\frac{s}{h}}^{0}\right)\\
 & = & \lim_{\epsilon\to0}\frac{1}{\epsilon}\left(f(\mbox{\ensuremath{\alpha}}x_{\frac{s}{h}}^{\epsilon}+(1-\alpha)x_{\frac{s}{h}-1})-f(\mbox{\ensuremath{\alpha}}x_{\frac{s}{h}}^{0}+(1-\alpha)x_{\frac{s}{h}-1})\right)\,h+\alpha\delta_{\frac{s}{h},\frac{s}{h}}+(1-\alpha)\delta_{\frac{s}{h}-1,\frac{s}{h}}\\
 & \stackrel{h\to0}{=} & \alpha,
\end{eqnarray*}
because the contribution of the deterministic evolution vanishes due
to the factor $h$. So for $\alpha=0$ (Ito convention) we have $\chi(s,s)=0$,
for $\alpha=\frac{1}{2}$ (Stratonovich) we have $\chi(s,s)=\frac{1}{2}$.
The Ito-convention is advantageous in this respect, because it leads
to vanishing contributions in Feynman diagrams  with response functions
at equal time points \citep{Chow15}.

We also observe that the initial condition contributes a term $-a\delta_{l,0}$.
Consequently, the initial condition can alternatively be included
by setting $a=0$ and instead calculate all moments from the generating
functional $Z[j,\tilde{j}-a\delta]$ instead of $Z[j,\tilde{j}]$.
In the following we will therefore skip the explicit term ensuring
the proper initial condition as it can be inserted by choosing the
proper value for the source $\tilde{j}$. See also \citep[Sec. 5.5]{hertz2016_arxiv}.

For the important special case of Gaussian white noise \eqref{eq:GWN_OM},
the generating functional, including the source field $\tilde{j}$
coupling to the response field, takes the form

\begin{align}
Z[j,\tilde{j}] & =\int\D x\,\int\D_{2\pi i}\tilde{x}\,\exp\left(\tilde{x}^{\T}(\partial_{t}x-f(x))+\frac{D}{2}\tx^{\T}\tx+j^{\T}x+\tilde{j}^{\T}\tilde{x}\right),\label{eq:MSR_GWN}
\end{align}
where we again used the definition of the inner product \eqref{eq:inner_product}.

\section{Dynamic mean-field theory for random networks\label{sec:Sompolinsky-Crisanti-Sommers-theory}}

Systems with many interacting degrees of freedom present a central
quest in physics. While disordered equilibrium systems show fascinating
properties such as the spin-glass transition \citep{Parisi80_1101,Sompolinsky81},
new collective phenomena arise in non-equilibrium systems: Large random
networks of neuron-like units can exhibit chaotic dynamics \citep{Sompolinsky88_259,Vreeswijk96,Monteforte10_268104}
with important functional consequences. In particular, information
processing capabilities show optimal performance close to the onset
of chaos \citep{Legenstein07_323,Sussillo09_544,Toyoizumi11_051908}.

Until today, the seminal work by \citet{Sompolinsky88_259} has a
lasting influence on the research field of random recurrent neural
networks, presenting a solvable random network model with deterministic
continuous-time dynamics that admits a calculation of the transition
to a chaotic regime and a characterization of chaos by means of Lyapunov
exponents. Many subsequent studies have built on top of this work
\citep{rajan10_011903,Hermann12_018702,wainrib13_118101,Aljadeff15_088101,Kadmon15_041030,Goedeke16_arxiv}.

The presentation in the original work \citep{Sompolinsky88_259},
published in \emph{Physical Review Letters}, summarizes the main steps
of the derivations and the most important results. In this chapter
we would like to show the formal calculations that, in our view, reproduce
the most important results. In lack of an extended version of the
original work, we do not know if the calculations by the original
authors are identical to the presentation here. However, we hope that
the didactic presentation given here may be helpful to provide an
easier access to the original work.

Possible errors in this document should not be attributed to the original
authors, but to the authors of this manuscript. In deriving the theory,
we also present a recent extension of the model to stochastic dynamics
due to additive uncorrelated Gaussian white noise \citep{Goedeke16_arxiv}.
The original results of \citep{Sompolinsky88_259} are obtained by
setting the noise amplitude $D=0$ in all expressions.

\subsection{Definition of the model and generating functional\label{sec:generating_functional}}

We study the coupled set of first order stochastic differential equations
\begin{align}
d\bx(t)+\bx(t)\,dt & =\bJ\phi(\bx(t))\,dt+d\bW(t),\label{eq:diffeq_motion}
\end{align}
where 
\begin{align}
J_{ij} & \sim\begin{cases}
\N(0,\frac{g^{2}}{N})\,\text{i.i.d.} & \text{for }i\neq j\\
0 & \text{for }i=j
\end{cases}\label{eq:connectivity_distribution}
\end{align}
are i.i.d. Gaussian random couplings, $\phi$ is a non-linear gain
function applied element-wise, the $dW_{i}$ are pairwise uncorrelated
Wiener processes with $\langle dW{}_{i}^{2}(t)\rangle=D\,dt$. For
concreteness we will use 
\begin{align}
\phi(x) & =\tanh(x),\label{eq:gain_function}
\end{align}
as in the original work \citep{Sompolinsky88_259}.

We formulate the problem in terms of a generating functional from
which we can derive all moments of the activity as well as response
functions. Introducing the notation $\tbx^{T}\bx=\sum_{i}\int\,\tx_{i}(t)x_{i}(t)\,dt$,
we obtain the moment-generating functional (cf. eq. (20)) 

\begin{align}
Z[\bl,\tbj](\bJ) & =\int\D\bx\int\D\tbx\,\exp\Big(S_{0}[\bx,\tbx]-\tbx^{\T}\bJ\phi\left(\bx\right)+\bl^{\T}\bx+\tbj^{\T}\tbx\Big)\nonumber \\
\text{with }S_{0}[\bx,\tbx] & =\tbx^{T}\left(\partial_{t}+1\right)\bx+\frac{D}{2}\tbx^{T}\tbx,\label{eq:def_S0}
\end{align}
where the measures are defined as $\int\D\bx=\lim_{M\to\infty}\Pi_{i=1}^{N}\Pi_{k=1}^{M}\int_{-\infty}^{\infty}dx_{i}^{k}$
and $\lim_{M\to\infty}\int\D\tbx=\Pi_{i=1}^{N}\Pi_{k=0}^{M-1}\int_{-i\infty}^{i\infty}\frac{d\tilde{x}_{i}^{k}}{2\pi i}$.
Here the superscript $k$ denotes the $k$-th time slice and we skip
the subscript $\D_{2\pi i}$, as introduced in \prettyref{eq:OM_pathint}
in \prettyref{sub:Onsager-Machlup-path-integral}, in the measure
of $\D\tbx$. The action $S_{0}$ is defined to contain all single
unit properties, therefore excluding the coupling term $-\tbx^{\T}\bJ\phi\left(\bx\right)$,
which is written explicitly.

\subsection{Average over the quenched disorder\label{sub:Disorder-average}}

The dynamics of \prettyref{eq:diffeq_motion} shows invariant features
independent of the actual realization of the couplings, only dependent
on their statistics, here parameterized by $g$. To capture these
properties that are generic to the ensemble of the models, we introduce
the averaged functional
\begin{eqnarray}
\bar{Z}[\bl,\tbj] & := & \langle Z[\bl,\tbj](\bJ)\rangle_{\bJ}\label{eq:disorder_averaged_Z}\\
 & = & \int\Pi_{ij}dJ_{ij}\,\mathcal{N}(0,\frac{g^{2}}{N},J_{ij})\,Z[\bl,\tbj](\bJ).\nonumber 
\end{eqnarray}
We use that the coupling term $\exp(-\sum_{i\neq j}J_{ij}\int\tilde{x}_{i}(t)\phi(x_{j}(t))\,dt)$
in \eqref{eq:def_S0} factorizes into $\Pi_{i\neq j}\exp(-J_{ij}\int\tilde{x}_{i}(t)\phi(x_{j}(t))\,dt)$
as does the distribution over the couplings (due to $J_{ij}$ being
independently distributed). We make use of the couplings appearing
linearly in the action and complete the square (in each $J_{ij}$
separately) to obtain for $i\neq j$
\begin{eqnarray}
 &  & \int dJ_{ij}\mathcal{N}(0,\frac{g^{2}}{N},J_{ij})\,\exp\left(-J_{ij}\int\tilde{x}_{i}(t)\phi(x_{j}(t))\,dt\right)\label{eq:completion_of_square}\\
 & = & \exp\left(\frac{g^{2}}{2N}\left(\int\tilde{x}_{i}(t)\phi(x_{j}(t))\,dt\right)^{2}\right).\nonumber 
\end{eqnarray}
We reorganize the last term including the sum $\sum_{i\neq j}$ as
\begin{eqnarray*}
 &  & \exp\left(\frac{g^{2}}{2N}\sum_{i\neq j}\left(\int\tilde{x}_{i}(t)\phi(x_{j}(t))\,dt\right)^{2}\right)\\
 & = & \exp\left(\frac{g^{2}}{2N}\sum_{i\neq j}\int\int\tilde{x}_{i}(t)\phi(x_{j}(t))\,\tilde{x}_{i}(t^{\prime})\phi(x_{j}(t^{\prime}))\,dt\,dt^{\prime}\right)\\
 & = & \exp\left(\frac{1}{2}\int\int\left(\sum_{i}\tilde{x}_{i}(t)\tilde{x}_{i}(t^{\prime})\right)\,\left(\frac{g^{2}}{N}\sum_{j}\phi(x_{j}(t))\phi(x_{j}(t^{\prime}))\right)\,dt\,dt^{\prime}\right)\times\\
 &  & \exp\left(-\frac{g^{2}}{2N}\,\int\int\sum_{i}\tilde{x}_{i}(t)\tilde{x}_{i}(t^{\prime})\phi(x_{i}(t))\phi(x_{i}(t^{\prime}))\,dt\,dt^{\prime}\right),
\end{eqnarray*}
where we used $\left(\int f(t)dt\right)^{2}=\int\int f(t)f(t^{\prime})\,dt\,dt^{\prime}$
in the first step and $\sum_{ij}x_{i}y_{j}=\sum_{i}x_{i}\sum_{j}y_{j}$
in the second. The last term is the diagonal element that is to be
taken out of the double sum. It is a correction of order $N^{-1}$
and will be neglected in the following. The disorder-averaged generating
functional \prettyref{eq:disorder_averaged_Z} therefore takes the
form
\begin{eqnarray}
\bar{Z}[\bl,\tbj] & = & \int\D\bx\int\D\tbx\,\exp\Big(S_{0}[\bx,\tbx]+\bl^{\T}\bx+\tbj^{\T}\tbx\Big)\times\label{eq:Zbar_pre}\\
 &  & \times\exp\Big(\frac{1}{2}\int_{-\infty}^{\infty}\int_{-\infty}^{\infty}\left(\sum_{i}\tilde{x}_{i}(t)\tilde{x}_{i}(t^{\prime})\right)\,\underbrace{\left(\frac{g^{2}}{N}\sum_{j}\phi(x_{j}(t))\phi(x_{j}(t^{\prime}))\right)}_{=:Q_{1}(t,t^{\prime})}\,dt\,dt^{\prime}\Big).\nonumber 
\end{eqnarray}
The coupling term in the last line contains quantities that depend
on four fields. We now aim to decouple these terms into terms of products
of pairs of fields. The aim is to make use of the central limit theorem,
namely that the quantity $Q_{1}$ indicated by the curly braces in
\prettyref{eq:Zbar_pre} is a superposition of a large ($N$) number
of (weakly correlated) contributions, which will hence approach a
Gaussian distribution. The outcome of the saddle point approximation
to lowest order will be the replacement of $Q_{1}$ by its expectation
value, as we will see in the following steps. We define

\begin{align}
Q_{1}(t,s):= & \frac{g^{2}}{N}\sum_{j}\phi(x_{j}(t))\phi(x_{j}(s))\label{eq:def_Q1}
\end{align}
and enforce this condition by inserting the Dirac-$\delta$ functional
\begin{align}
 & \delta[-\frac{N}{g^{2}}Q_{1}(s,t)+\sum_{j}\phi(x_{j}(s))\,\phi(x_{j}(t))]\label{eq:Hubbard_Stratonovich}\\
= & \int\D Q_{2}\,\exp\left(\iint\,Q_{2}(s,t)\left[-\frac{N}{g^{2}}Q_{1}(s,t)+\sum_{j}\,\phi(x_{j}(s))\,\phi(x_{j}(t))\right]\,ds\,dt\right).\nonumber 
\end{align}
We here note that as for the response field, the field $Q_{2}\in i\mathbb{R}$
is purely imaginary due to the Fourier representation \prettyref{eq:Fourier_delta}
of the $\delta$.

We aim at a set of self-consistent equations for the auxiliary fields.
We therefore introduce one source term for each of the fields to be
determined. Extending our notation by defining $Q_{1}^{\T}Q_{2}:=\iint\,Q_{1}(s,t)\,Q_{2}(s,t)\,ds\,dt$
and $\tx^{\T}Q_{1}\tx:=\iint\,\tx(s)\,Q_{1}(s,t)\,\tx(t)\,ds\,dt$
we hence rewrite \prettyref{eq:Zbar_pre} as

\begin{eqnarray}
\bar{Z}[j,\tj] & = & \int\D Q_{1}\int\D Q_{2}\,\exp\left(-\frac{N}{g^{2}}Q_{1}^{T}Q_{2}+N\,\ln\,Z[Q_{1},Q_{2}]+j^{\T}Q_{1}+\tj^{\T}Q_{2}\right)\label{eq:Zbar}\\
Z[Q_{1},Q_{2}] & = & \int\D x\int\D\tx\,\exp\Big(S_{0}[x,\tx]+\nonumber \\
 &  & +\frac{1}{2}\tx^{\T}Q_{1}\tx+\phi(x)^{\T}Q_{2}\phi(x)\Big),\nonumber 
\end{eqnarray}
where the integral measures $\D Q_{1,2}$ must be defined suitably.
In writing $N\,\ln\,Z[Q_{1},Q_{2}]$ we have used that the auxiliary
fields couple only to sums of fields $\sum_{i}\phi^{2}(x_{i})$ and
$\sum_{i}\tx_{i}^{2}$, so that the generating functional for the
fields $\bx$ and $\tbx$ factorizes into a product of $N$ factors
$Z[Q_{1},Q_{2}]$. The latter only contains functional integrals over
the two scalar fields $x$, $\tx$. This shows that we have reduced
the problem of $N$ interacting units to that of a single unit exposed
to a set of external fields $Q_{1}$ and $Q_{2}$.

The remaining problem can be considered a field theory for the auxiliary
fields $Q_{1}$ and $Q_{2}$. The form \prettyref{eq:Zbar} clearly
exposes the $N$ dependence of the action for these latter fields
in \prettyref{eq:Zbar}: It is of the form $\int dQ\exp(Nf(Q))\,dQ$,
which, for large $N$, suggests a saddle point approximation.

\begin{figure}
\begin{centering}
\includegraphics[scale=0.5]{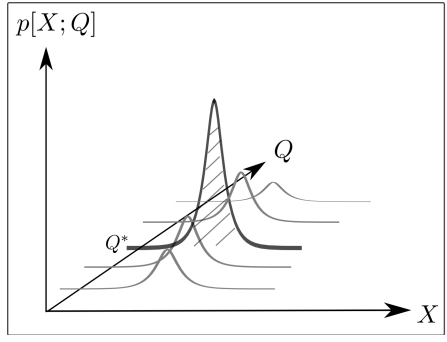}
\par\end{centering}
\caption{\textbf{Finding saddle point by maximizing contribution to probability:}
The contribution to the overall probability mass depends on the value
of the parameter $Q$, i.e. we seek to maximize $b[Q]:=\int\protect\D x\,p[\protect\bx;Q]$
\eqref{eq:def_b}. The point at which the maximum is attained is denoted
as $Q^{\ast}$, the value $b[Q^{\ast}]$ is indicated by the hatched
area.\label{fig:intuition_saddle_point_max}}
\end{figure}

In the saddle point approximation \citep{Sompolinsky82_6860} we seek
the stationary point of the action determined by 
\begin{eqnarray}
0=\frac{\delta S[Q_{1},Q_{2}]}{\delta Q_{\{1,2\}}}=\frac{\delta}{\delta Q_{\{1,2\}}}\left(-\frac{N}{g^{2}}Q_{1}^{T}Q_{2}+N\,\ln Z[Q_{1},Q_{2}]\right) & = & 0.\label{eq:def_saddle_Q1_Q2}
\end{eqnarray}
We here set the value for the source fields $j=\tj=0$ to zero. This
corresponds to finding the point in the space $(Q_{1},Q_{2})$ which
provides the dominant contribution to the probability mass. This can
be seen by writing the probability functional as $p[\bx]=\iint\D Q_{1}\D Q_{2}\,p[\bx;Q_{1},Q_{2}]$
with 
\begin{align}
p[\bx;Q_{1},Q_{2}] & =\exp\left(-\frac{N}{g^{2}}Q_{1}^{T}Q_{2}+\sum_{i}\ln\int\D\tx\,\exp\Big(S_{0}[x_{i},\tx]+\frac{1}{2}\tx^{\T}Q_{1}\tx+\phi(x_{i})^{\T}Q_{2}\phi(x_{i})\Big)\right)\nonumber \\
b[Q_{1},Q_{2}] & :=\int\D\bx\,p[\bx;Q_{1},Q_{2}],\label{eq:def_b}
\end{align}
where we defined $b[Q_{1},Q_{2}]$ as the contribution to the entire
probability mass for a given value of the auxiliary fields $Q_{1},Q_{2}$.
Maximizing $b$ therefore amounts to the condition \prettyref{eq:def_saddle_Q1_Q2},
illustrated in \prettyref{fig:intuition_saddle_point_max}. We here
used the convexity of the exponential function. 

A more formal argument to obtain \prettyref{eq:def_saddle_Q1_Q2}
proceeds by introducing the Legendre-Fenchel transform of $\ln\bar{Z}$
as 
\begin{eqnarray*}
\Gamma(q_{1},q_{2}) & := & \sup_{j,\tj}\,j^{\T}q_{1}-\ln\bar{Z}[j,\tj],
\end{eqnarray*}
called the \textbf{vertex generating functional} or \textbf{effective
action} \citep{ZinnJustin96,NegeleOrland98}. It holds that $\frac{\delta\Gamma}{\delta q_{1}}=j$
and $\frac{\delta\Gamma}{\delta q_{2}}=\tj$, called \textbf{equations
of state}. The leading order mean-field approximation amounts to
the approximation $\Gamma[q_{1},q_{2}]\simeq-S[q_{1},q_{2}]$. The
equations of state, for vanishing sources $j=\tj=0$, therefore yield
the saddle point equations 
\begin{align*}
0=\frac{\delta\Gamma}{\delta q_{1}} & =-\frac{\delta S}{\delta q_{1}}\\
0=\frac{\delta\Gamma}{\delta q_{2}} & =-\frac{\delta S}{\delta q_{2}},
\end{align*}
identical to \prettyref{eq:def_saddle_Q1_Q2}. This more formal view
has the advantage of being straight forwardly extendable to loopwise
corrections.

The functional derivative in the stationarity condition \prettyref{eq:def_saddle_Q1_Q2}
applied to $\ln Z[Q_{1},Q_{2}]$ produces an expectation value with
respect to the distribution \prettyref{eq:def_b}: the fields $Q_{1}$
and $Q_{2}$ here act as sources. This yields the set of two equations
\begin{eqnarray}
0=-\frac{N}{g^{2}}\,Q_{1}^{\ast}(t,t^{\prime})+\frac{N}{Z}\,\left.\frac{\delta Z[Q_{1},Q_{2}]}{\delta Q_{2}(s,t)}\right|_{Q^{\ast}} & \leftrightarrow & Q_{1}^{\ast}(s,t)=g^{2}\left\langle \phi(x(s))\phi(x(t))\right\rangle _{Q^{\ast}}=:g^{2}C_{\phi(x)\phi(x)}(t,t^{\prime})\label{eq:saddle_Q1_Q2}\\
0=-\frac{N}{g^{2}}\,Q_{2}^{\ast}(t,t^{\prime})+\frac{N}{Z}\,\left.\frac{\delta Z[Q_{1},Q_{2}]}{\delta Q_{1}(s,t)}\right|_{Q^{\ast}} & \leftrightarrow & Q_{2}^{\ast}(s,t)=\frac{g^{2}}{2}\langle\tilde{x}(s)\tilde{x}(t)\rangle_{Q^{\ast}}=0,\nonumber 
\end{eqnarray}
where we defined the average autocorrelation function $C_{\phi(x)\phi(x)}(t,t^{\prime})$
of the non-linearly transformed activity of the units. The second
saddle point $Q_{2}^{\ast}=0$ vanishes, as it would otherwise alter
the normalization of the generating functional through mixing of retarded
and non-retarded time derivatives which then yield acausal response
functions \citep{Sompolinsky82_6860}. 

The expectation values $\langle\rangle_{Q^{\ast}}$ appearing in \eqref{eq:saddle_Q1_Q2}
must be computed self-consistently, since the values of the saddle
points, by \prettyref{eq:Zbar}, influence the statistics of the fields
$\bx$ and $\tbx$, which in turn determines the functions $Q_{1}^{\ast}$
and $Q_{2}^{\ast}$ by \eqref{eq:saddle_Q1_Q2}.

\begin{figure}
\begin{centering}
\includegraphics[scale=0.5]{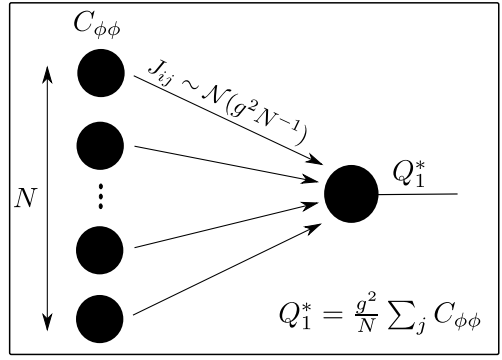}
\par\end{centering}
\caption{\textbf{Interpretation of the saddle point value $Q_{1}^{\ast}$ given
by eq. \prettyref{eq:saddle_Q1_Q2}:} The summed covariances $C_{\phi\phi}$
received by a neuron in the network, weighted by the synaptic couplings
$J_{ij}$, which have Gaussian statistics with variance $g^{2}N^{-1}$.\label{fig:intuition_saddle_point}}
\end{figure}

Inserting the saddle point solution into the generating functional
\eqref{eq:Zbar} we get
\begin{eqnarray}
\bar{Z}^{\ast} & \propto & \int\D x\int\D x\,\exp\,\Big(S_{0}[x,\tx]+\frac{g^{2}}{2}\tx^{\T}C_{\phi(x)\phi(x)}\tx\Big).\label{eq:Z_bar_star}
\end{eqnarray}
As the saddle points only couple to the sums of fields, the action
has the important property that it decomposes into a sum of actions
for individual, non-interacting units that feel a common field with
self-consistently determined statistics, characterized by its second
cumulant $C_{\phi(x)\phi(x)}$. Hence the saddle-point approximation
reduces the network to $N$ non-interacting units, or, equivalently,
a single unit system. The second term in \prettyref{eq:Z_bar_star}
is a Gaussian noise with a two point correlation function $C_{\phi(x)\phi(x)}(t,t^{\prime})$.
The physical interpretation is the noisy signal each unit receives
due to the input from the other $N$ units. Its autocorrelation function
is given by the summed autocorrelation functions of the output activities
$\phi(x_{i}(t))$ weighted by $g^{2}N^{-1}$, which incorporates
the Gaussian statistics of the couplings. This intuitive picture is
shown in \prettyref{fig:intuition_saddle_point}.

The interpretation of the noise can be appreciated by explicitly considering
the moment generating functional of a Gaussian noise with a given
autocorrelation function $C(t,t^{\prime})$, which leads to the cumulant
generating functional $\ln Z_{\zeta}[\tilde{x}]$ that appears in
the exponent of \eqref{eq:Z_bar_star} and has the form
\begin{eqnarray*}
\ln\,Z_{\zeta}[\tilde{x}] & = & \ln\langle\exp\left(\int\tx(t)\,\zeta(t)\,dt\right)\rangle\\
 & = & \frac{1}{2}\int_{-\infty}^{\infty}\int_{-\infty}^{\infty}\tilde{x}(t)\,C(t,t^{\prime})\,\tilde{x}(t^{\prime})\,dt\,dt^{\prime}\\
 & = & \frac{1}{2}\tx^{\T}\,C\,\tx.
\end{eqnarray*}
Note that the effective noise term only has a non-vanishing second
cumulant. This means the effective noise is Gaussian, as the cumulant
generating function is quadratic. It couples pairs of time points
that are correlated.

This is the starting point in \citep[ eq. (3)]{Sompolinsky88_259},
stating that the effective mean-field dynamics of the network is given
by that of a single unit
\begin{eqnarray}
(\partial_{t}+1)\,x(t) & = & \eta(t)\label{eq:effective_mf_Langevin}
\end{eqnarray}
driven by a Gaussian noise $\eta=\zeta+\frac{dW}{dt}$ with autocorrelation
$\langle\eta(t)\eta(s)\rangle=g^{2}\,C_{\phi(x)\phi(x)}(t,s)+D\delta(t-s)$.
In the cited paper the white noise term $\propto D$ is absent, though.

We multiply the equation \prettyref{eq:effective_mf_Langevin} for
time points $t$ and $s$ and take the expectation value with respect
to the noise $\eta$ on both sides, which leads to
\begin{align}
\left(\partial_{t}+1\right)\left(\partial_{s}+1\right)C_{xx}(t,s) & =g^{2}\,C_{\phi(x)\phi(x)}(t,s)+D\delta(t-s),\label{eq:cov_xx_diffeq}
\end{align}
where we defined the covariance function of the activities $C_{xx}(t,s):=\langle x(t)x(s)\rangle$.
In the next section we will rewrite this equation into an equation
of a particle in a potential.\textbf{}

\subsection{Stationary statistics: Self-consistent autocorrelation of as motion
of a particle in a potential\label{sub:particle_motion}}

We are now interested in the stationary statistics of the system,
i.e. $C_{xx}(t,s)=:c(t-s)$. The inhomogeneity in \eqref{eq:cov_xx_diffeq}
is then also time-translation invariant, $C_{\phi(x)\phi(x)}(t+\tau,t)$
is only a function of $\tau$. Therefore the differential operator
$\left(\partial_{t}+1\right)\left(\partial_{s}+1\right)c(t-s)$, with
$\tau=t-s$, simplifies to $(-\partial_{\tau}^{2}+1)\,c(\tau)$ so
we get
\begin{eqnarray}
(-\partial_{\tau}^{2}+1)\,c(\tau) & = & g^{2}\,C_{\phi(x)\phi(x)}(t+\tau,t)+D\,\delta(\tau).\label{eq:diffeq_auto}
\end{eqnarray}
Once \prettyref{eq:diffeq_auto} is solved, we know the covariance
function $c(\tau)$ between two time points $\tau$ apart as well
as the variance $c(0)=:c_{0}$. Since by the saddle point approximation
in \prettyref{sub:Disorder-average} the expression \prettyref{eq:Z_bar_star}
is the generating functional of a Gaussian theory, the $x_{i}$ are
zero mean Gaussian random variables. Consequently the second moment
completely determines the distribution. We can therefore obtain $C_{\phi(x)\phi(x)}(t,s)=g^{2}f_{\phi}(c(\tau),c_{0})$
with 
\begin{align}
f_{u}(c,c_{0}) & =\iint\,u\Bigg(\sqrt{c_{0}-\frac{c^{2}}{c_{0}}}\,z_{1}+\tfrac{c}{\sqrt{c_{0}}}\,z_{2}\Bigg)u\Bigg(\sqrt{c_{0}}\,z_{2}\Bigg)\,Dz_{1}Dz_{2}\label{eq:def_f}
\end{align}
with the Gaussian integration measure $Dz=\exp(-z^{2}/2)/\sqrt{2\pi}\,dz$
and for a function $u(x)$. Here, the two different arguments of $u(x)$
are by construction Gaussian with zero mean, variance $c(0)=c_{0}$,
and covariance $c(\tau)$. Note that \prettyref{eq:def_f} reduces
to one-dimensional integrals for $f_{u}(c_{0},c_{0})=\langle u(x)^{2}\rangle$
and $f_{u}(0,c_{0})=\langle u(x)\rangle^{2}$, where $x$ has zero
mean and variance $c_{0}$.

We note that $f_{u}(c(\tau),c_{0})$ in \prettyref{eq:def_f} only
depends on $\tau$ through $c(\tau)$. We can therefore obtain it
from the ``potential'' $g^{2}f_{\Phi}(c(\tau),c_{0})$ by 
\begin{eqnarray}
C_{\phi(x)\phi(x)}(t+\tau,t) & =: & \frac{\partial}{\partial c}\,g^{2}f_{\Phi}(c(\tau),c_{0})\label{eq:def_potential_v}
\end{eqnarray}
where $\Phi$ is the integral of $\phi$, i.e. $\Phi(x)=\int_{0}^{x}\phi(x)\,dx=\ln\cosh(x)$.
The property $\frac{\partial}{\partial c}\,g^{2}f_{\Phi}(c,c_{0})=\,g^{2}f_{\Phi^{\prime}}(c(\tau),c_{0})$
(Price's theorem \citep{PapoulisProb}) is shown in the supplementary
calculation in Section IIIA. Note that the representation in \prettyref{eq:def_f}
differs from the one used in \citep[eq. (7)]{Sompolinsky88_259}.
The expression used here is also valid for negative $c(\tau)$ in
contrast to the original formulation. We can therefore express the
differential equation for the autocorrelation with the definition
of the potential $V$ 
\begin{eqnarray}
V(c;c_{0}) & := & -\frac{1}{2}c^{2}+g^{2}f_{\Phi}(c(\tau),c_{0})-g^{2}f_{\Phi}(0,c_{0}),\label{eq:def_potential_V}
\end{eqnarray}
where the subtraction of the last constant term is an arbitrary choice
that ensures that $V(0;c_{0})=0$. The equation of motion \prettyref{eq:diffeq_auto}
therefore takes the form

\begin{eqnarray}
\partial_{\tau}^{2}\,c(\tau) & = & -V^{\prime}(c(\tau);c_{0})-D\,\delta(\tau),\label{eq:eq_motion_cxx}
\end{eqnarray}
so it describes the motion of a particle in a (self-consistent) potential
$V$ with derivative $V^{\prime}=\frac{\partial}{\partial c}V$. The
$\delta$-distribution on the right hand side causes a jump in the
velocity that changes from $\frac{D}{2}$ to $-\frac{D}{2}$ at $\tau=0$,
because $c$ is symmetric ($c(\tau)=c(-\tau)$) and hence $\dot{c}(\tau)=-\dot{c}(-\tau)$
and moreover the term $-V^{\prime}(c(\tau);c_{0})$ does not contribute
to the kink. The equation must be solved self-consistently, as the
initial value $c_{0}$ determines the effective potential $V(\cdot,c_{0})$
via \eqref{eq:def_potential_V}. The second argument $c_{0}$ indicates
this dependence.

\begin{figure}
\begin{centering}
\includegraphics[scale=2]{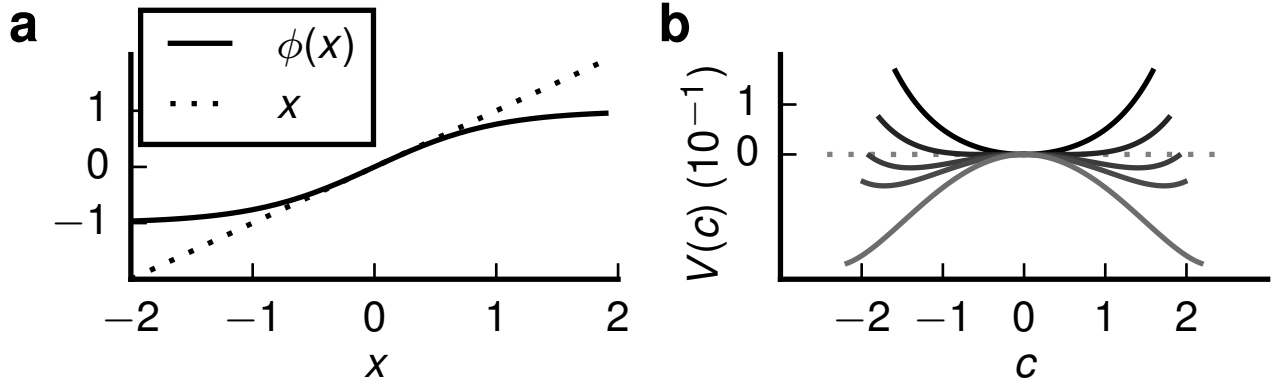}
\par\end{centering}
\caption{\textbf{Effective potential for the noise-less case $D=0$. a }The
gain function $\phi(x)=\tanh(x)$ close to the origin has unit slope.
Consequently, the integral of the gain function $\Phi(x)=\ln\cosh(x)$
close to origin has the same curvature as the parabola $\frac{1}{2}x^{2}$.
\textbf{b }Self-consistent potential for $g=2$ and different values
of $c_{0}=1.6,1.8,1.924,2,2.2$ (from black to light gray). The horizontal
gray dotted line indicates the identical levels of initial and finial
potential energy for the self-consistent solution $V(c_{0};c_{0})=0$,
corresponding to the initial value that leads to a monotonously decreasing
autocovariance function that vanishes for $\tau\to\infty$.\label{fig:effective_potential}}
\end{figure}

The gain function $\phi(x)=\tanh(x)$ is shown in \prettyref{fig:effective_potential}a,
while \prettyref{fig:effective_potential}b shows the self-consistent
potential for the noiseless case $D=0$.

The potential is formed by the interplay of two opposing terms. The
downward bend is due to $-\frac{1}{2}c^{2}$. The term $g^{2}f_{\Phi}(c;c_{0})$
is bent upwards. We get an estimate of this term from its derivative
$g^{2}f_{\phi}(c,c_{0})$: Since $\phi(x)$ has unit slope at $x=0$
(see \prettyref{fig:effective_potential}a), for small amplitudes
$c_{0}$ the fluctuations are in the linear part of $\phi$, so $g^{2}f_{\phi}(c,c_{0})\simeq g^{2}c$
for all $c\le c_{0}$. Consequently, the potential $g^{2}f_{\Phi}(c,c_{0})=\int_{0}^{c}g^{2}f_{\phi}(c^{\prime},c_{0})\,dc^{\prime}\stackrel{c<c_{0}\ll1}{\simeq}g^{2}\frac{1}{2}c^{2}$
has a positive curvature at $c=0$.

For $g<1$, the parabolic part dominates for all $c_{0}$, so that
the potential is bent downwards and the only bounded solution in the
noiseless case $D=0$ of \prettyref{eq:eq_motion_cxx} is the vanishing
solution $c(t)\equiv0$.

For $D>0$, the particle may start at some point $c_{0}>0$ and, due
to its initial velocity, reach the point $c(\infty)=0$. Any physically
reasonable solution must be bounded. In this setting, the only possibility
is a solution that starts at a position $c_{0}>0$ with the same initial
energy $V(c_{0};c_{0})+E_{\mathrm{kin}}^{0}$ as the final potential
energy $V(0;c_{0})=0$ at $c=0$. The initial kinetic energy is given
by the initial velocity $\dot{c}(0+)=-\frac{D}{2}$ as $E_{\mathrm{kin}}^{(0)}=\frac{1}{2}\dot{c}(0+)^{2}=\frac{D^{2}}{8}$.
This condition ensures that the particle, starting at $\tau=0$ at
the value $c_{0}$ for $\tau\to\infty$ reaches the local maximum
of the potential at $c=0$; the covariance function decays from $c_{0}$
to zero.

For $g>1$, the term $g^{2}f_{\Phi}(c;c_{0})$ can start to dominate
the curvature close to $c\simeq0$: the potential in \prettyref{fig:effective_potential}b
is bent upwards for small $c_{0}$. For increasing $c_{0}$, the fluctuations
successively reach the shallower parts of $\phi$, hence the slope
of $g^{2}f_{\phi}(c,c_{0})$ diminishes, as does the curvature of
its integral, $g^{2}f_{\Phi}(c;c_{0})$. With increasing $c_{0}$,
the curvature of the potential at $c=0$ therefore changes from positive
to negative.

In the intermediate regime, the potential assumes a double well shape.
Several solutions exist in this case. One can show that the only stable
solution is the one that decays to $0$ for $\tau\to\infty$ \citep{Sompolinsky88_259}.
In the presence of noise $D>0$ this assertion is clear due to the
decorrelating effect of the noise, but it remains true also in the
noiseless case.

By the argument of energy conservation, the corresponding value $c_{0}$
can be found numerically as the root of 
\begin{eqnarray}
V(c_{0};c_{0})+E_{\mathrm{kin}}^{(0)} & \stackrel{!}{=} & 0\label{eq:initial_c0}\\
E_{\mathrm{kin}}^{(0)} & = & \frac{D^{2}}{8},\nonumber 
\end{eqnarray}
for example with a simple bisectioning algorithm.

\begin{figure}
\begin{centering}
\includegraphics[scale=2]{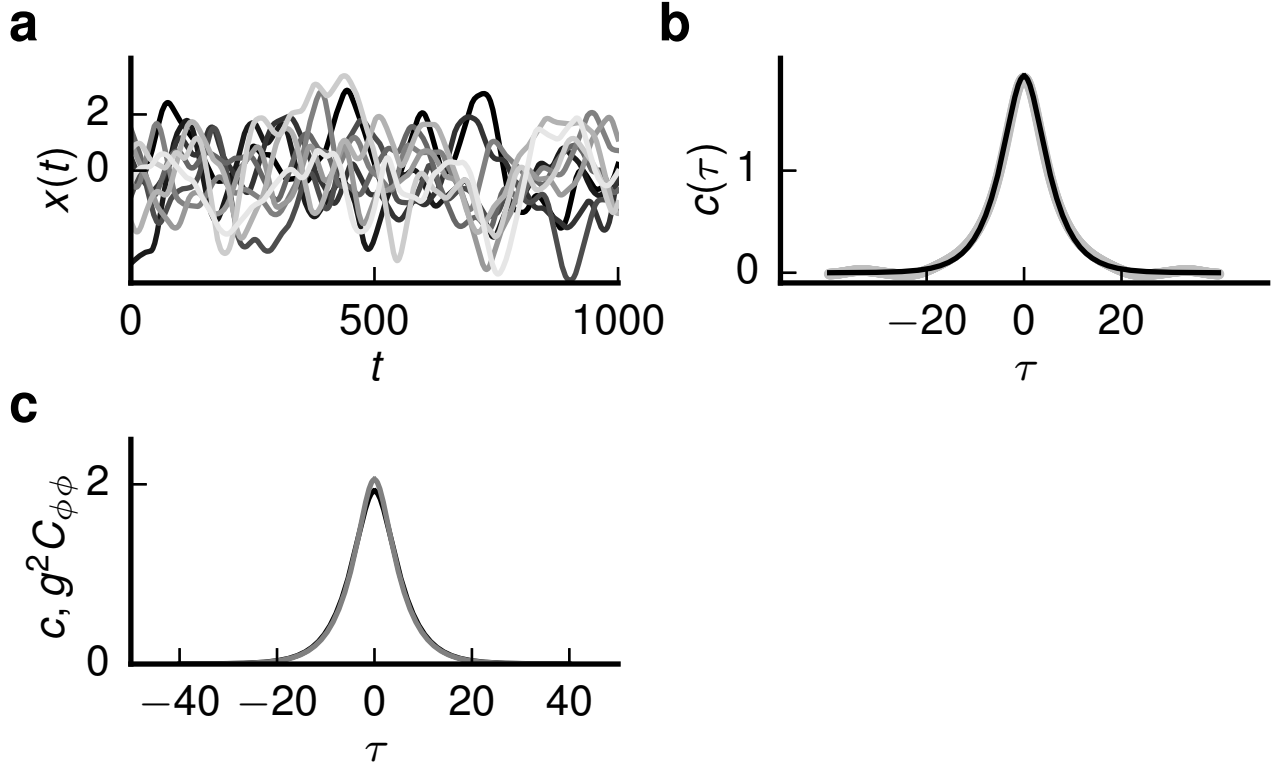}
\par\end{centering}
\caption{\textbf{Self-consistent autocovariance function from dynamic mean-field
theory in the noise-less case.} Random network of $5000$ Gaussian
coupled units with with $g=2$ and vanishing noise $D=0$. \textbf{a}
Activity of the first $10$ units as function of time. \textbf{b}
Self-consistent solution of covariance $c(\tau)$ (black) and result
from simulation (gray). The theoretical result is obtained by first
solving \prettyref{eq:initial_c0} for the initial value $c_{0}$
and then integrating \prettyref{eq:acf_system}. \textbf{c }Self-consistent
solution (black) as in b and $C_{\phi\phi}(\tau)=g^{2}f_{\phi}(c(\tau),c_{0})$
given by \prettyref{eq:def_potential_v} (gray). Duration of simulation
$T=1000$ time steps with resolution $h=0.1$ each. Integration of
\prettyref{eq:diffeq_motion} by forward Euler method.\label{fig:Self-consistent-solution-corr}}
\end{figure}

The corresponding shape of the autocovariance function then follows
a straight forward integration of the differential equation \prettyref{eq:eq_motion_cxx}.
Rewriting the second order differential equation into a coupled set
of first order equations, introducing $\partial_{\tau}c=:y$, we get
for $\tau>0$
\begin{eqnarray}
\partial_{\tau}\left(\begin{array}{c}
y(\tau)\\
c(\tau)
\end{array}\right) & = & \left(\begin{array}{c}
c-g^{2}f_{\phi}(c,c_{0})\\
y(\tau)
\end{array}\right)\label{eq:acf_system}\\
\text{with initial condition}\nonumber \\
\left(\begin{array}{c}
y(0)\\
c(0)
\end{array}\right) & = & \left(\begin{array}{c}
-\frac{D}{2}\\
c_{0}
\end{array}\right).\nonumber 
\end{eqnarray}
The solution of this equation in comparison to direct simulation is
shown in \prettyref{fig:Self-consistent-solution-corr}. Note that
the covariance function of the input to a unit, $C_{\phi\phi}(\tau)=g^{2}f_{\phi}(c(\tau),c_{0})$,
bares strong similarities to the autocorrelation $c$, shown in \prettyref{fig:Self-consistent-solution-corr}c:
The suppressive effect of the non-linear, saturating gain function
is compensated by the variance of the connectivity $g^{2}>1$, so
that a self-consistent solution is achieved.

\subsection{Assessing chaos by a pair of identical systems\label{sub:pair_of_systems}}

We now aim to study whether the dynamics is chaotic or not. To this
end, we consider a pair of identically prepared systems, in particular
with identical coupling matrix $\bJ$ and, for $D>0$, also the same
realization of the Gaussian noise. We distinguish the dynamical variables
$x^{\alpha}$ of the two systems by superscripts $\alpha\in\{1,2\}$.

\begin{figure}
\begin{centering}
\includegraphics[scale=2]{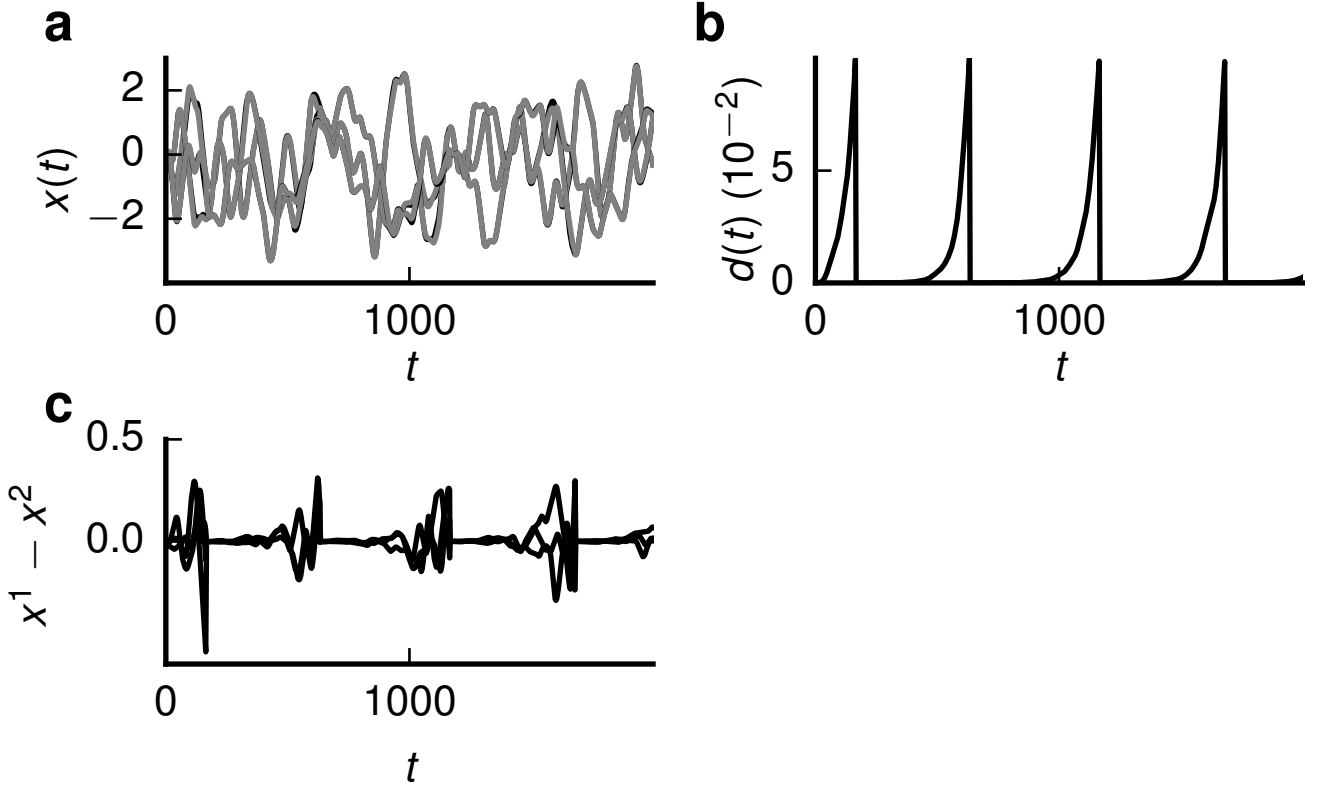}
\par\end{centering}
\caption{\textbf{Chaotic evolution. a }Dynamics of two systems starting at
similar initial conditions for chaotic case with $g=2$, $N=5000,$
$D=0.01$. Trajectories of three units shown for the unperturbed (black)
and the perturbed system (gray). \textbf{b }Absolute average squared
distance $d(t)$ given by \prettyref{eq:def_d_empirical} of the two
systems. \textbf{c }Difference $x_{1}-x_{2}$ for the first three
units. The second system is reset to the state of the first system
plus a small random displacement as soon as $d(t)>0.1$. Other parameters
as in \prettyref{fig:Self-consistent-solution-corr}. \label{fig:Chaotic-evolution-pair}}
\end{figure}

Let us briefly recall that the dynamical mean-field theory describes
empirical population-averaged quantities for a single network realization
(due to self-averaging). Hence, for large $N$ we expect that 
\begin{align*}
\frac{1}{N}\sum_{i=1}^{N}x_{i}^{\alpha}(t)x_{i}^{\beta}(s) & \simeq c^{\alpha\beta}(t,s)
\end{align*}
holds for most network realizations. To study the stability of the
dynamics with respect to perturbations of the initial conditions we
consider the population-averaged (mean-)squared distance between the
trajectories of the two copies of the network:
\begin{align}
\frac{1}{N}||x^{1}(t)-x^{2}(t)||^{2} & =\frac{1}{N}\sum_{i=1}^{N}\left(x_{i}^{1}(t)-x_{i}^{2}(t)\right)^{2}\label{eq:def_d_empirical}\\
 & =\frac{1}{N}\sum_{i=1}^{N}\left(x_{i}^{1}(t)\right)^{2}+\frac{1}{N}\sum_{i=1}^{N}\left(x_{i}^{2}(t)\right)^{2}-\frac{2}{N}\sum_{i=1}^{N}x_{i}^{1}(t)x_{i}^{2}(t)\nonumber \\
 & \simeq c^{11}(t,t)+c^{22}(t,t)-2c^{12}(t,t)\,.\nonumber 
\end{align}
This idea has also been employed in \citep{Derrida87_721}. Therefore,
we define the mean-field mean-squared distance between the two copies:
\begin{align}
d(t,s) & :=c^{11}(t,s)+c^{22}(t,s)-c^{12}(t,s)-c^{21}(t,s)\,,\label{eq:mean-squared-distance-1}
\end{align}
which gives for equal time arguments the actual mean-squared distance
$d(t):=d(t,t)\,$. Our goal is to find the temporal evolution of $d(t,s)\,$.
The time evolution of a pair of systems in the chaotic regime with
slightly different initial conditions is shown in \prettyref{fig:Chaotic-evolution-pair}.
Although the initial displacement between the two systems is drawn
independently for each of the four shown trials, the divergence of
$d(t)$ has a stereotypical form, which seems to be dominated by one
largest Lyapunov exponent. The aim of the remainder of this section
is to find this rate of divergence.

To derive an equation of motion for $d(t,s)$ it is again convenient
to define a generating functional that captures the joint statistics
of two systems and in addition allows averaging over the quenched
disorder \citep[see also ][Appendix 23, last remark]{ZinnJustin96}.

The generating functional is defined in analogy to the single system
\eqref{eq:def_S0}
\begin{align}
Z[\{\bl^{\alpha},\tbj^{\alpha}\}_{\alpha\in\{1,2\}}](\bJ) & =\Pi_{\alpha=1}^{2}\Big\{\int\D\bx^{\alpha}\int\D\tbx^{\alpha}\,\exp\left(\tbx^{\alpha\T}\left((\partial_{t}+1)\,\bx^{\alpha}-\sum_{j}\bJ\phi(\bx^{\alpha})\right)+\bl^{\alpha\T}\bx^{\alpha}+\tbj^{\alpha\T}\tbx^{\alpha}\right)\Big\}\times\nonumber \\
 & \times\exp\left(\frac{D}{2}\,(\tbx^{1}+\tbx^{2})^{T}(\tbx^{1}+\tbx^{2})\right)\Big\},\label{eq:Z_pair_pre}
\end{align}
where the last term is the moment generating functional due to the
white noise that is common to both subsystems. We note that the coupling
matrix $\bJ$ is the same in both subsystems as well. Using the notation
analogous to \eqref{eq:def_S0} and collecting the terms that affect
each individual subsystem in the first, the common term in the second
line, we get

\begin{align}
Z[\{\bl^{\alpha},\tbj^{\alpha}\}_{\alpha\in\{1,2\}}](\bJ) & =\Pi_{\alpha=1}^{2}\Big\{\int\D\bx^{\alpha}\int\D\tbx^{\alpha}\,\exp\Big(S_{0}[\bx^{\alpha},\tbx^{\alpha}]-\tbx^{\alpha\T}\bJ\phi\left(\bx^{\alpha}\right)+\bl^{\alpha\T}\bx^{\alpha}+\tbj^{\alpha\T}\tbx^{\alpha}\Big)\Big\}\nonumber \\
 & \times\exp\left(D\tbx^{1\T}\tbx^{2}\right).\label{eq:Z_pair}
\end{align}
Here the term in the last line appears due to the mixed product of
the response fields in \eqref{eq:Z_pair_pre}.

We will now perform the average over realizations in $\bJ$, as in
\prettyref{sub:Disorder-average} eq. \eqref{eq:completion_of_square}.
We therefore need to evaluate the Gaussian integral
\begin{align}
 & \int dJ_{ij}\mathcal{N}(0,\frac{g^{2}}{N},J_{ij})\,\exp\left(-J_{ij}\sum_{\alpha=1}^{2}\tilde{x}_{i}^{\alpha\T}\phi(x_{j}^{\alpha})\right)\nonumber \\
 & =\exp\left(\frac{g^{2}}{2N}\sum_{\alpha=1}^{2}\left(\tilde{x}_{i}^{\alpha\T}\phi(x_{j}^{\alpha})\right)^{2}\right)\nonumber \\
 & \times\exp\left(\frac{g^{2}}{N}\,\tilde{x}_{i}^{1\T}\phi(x_{j}^{1})\,\tilde{x}_{i}^{2\T}\phi(x_{j}^{2})\right).\label{eq:quenched_avg_pair}
\end{align}
Similar as for the Gaussian integral over the common noises that gave
rise to the coupling term between the two systems in the second line
of \eqref{eq:Z_pair}, we here obtain a coupling term between the
two systems, in addition to the terms that only include variables
of a single subsystem in the second last line. Note that the two coupling
terms are different in nature. The first, due to common noise, represents
common temporal fluctuations injected into both systems. The second
is static in its nature, as it arises from the two systems having
the same coupling $\bJ$ in each of their realizations that enter
the expectation value. The terms that only affect a single subsystem
are identical to those in \prettyref{eq:Zbar_pre}. We treat these
terms as before and here concentrate on the mixed terms, which we
rewrite (including the $\sum_{i\neq j}$ in \prettyref{eq:Z_pair}
and using our definition $\tilde{x}_{i}^{\alpha\T}\phi(x_{j}^{\alpha})=\int dt\,\tilde{x}_{i}^{\alpha}(t)\phi(x_{j}^{\alpha}(t))\,dt$)
as 
\begin{align}
 & \exp\Big(\frac{g^{2}}{N}\sum_{i\neq j}\,\tilde{x}_{i}^{1\T}\phi(x_{j}^{1})\,\tilde{x}_{i}^{2\T}\phi(x_{j}^{2})\Big)\label{eq:mixed_avg_pair}\\
= & \exp\Big(\iint\,\sum_{i}\tilde{x}_{i}^{1}(s)\tilde{x}_{i}^{2}(t)\underbrace{\frac{g^{2}}{N}\sum_{j}\phi(x_{j}^{1}(s))\,\phi(x_{j}^{2}(t))}_{=:T_{1}(s,t)}\,ds\,dt\Big)+O(N^{-1}),\nonumber 
\end{align}
where we included the self coupling term $i=j$, which is only a subleading
correction of order $N^{-1}$.

We now follow the steps in \prettyref{sub:Disorder-average} and introduce
three pairs of auxiliary variables. The pairs $Q_{1}^{\alpha},Q_{2}^{\alpha}$
are defined as before in \prettyref{eq:def_Q1} and \prettyref{eq:Hubbard_Stratonovich},
but for each subsystem, while the pair $T_{1},T_{2}$ decouples the
mixed term \prettyref{eq:mixed_avg_pair} by defining
\begin{align*}
T_{1}(s,t) & :=\frac{g^{2}}{N}\sum_{j}\phi(x_{j}^{1}(s))\,\phi(x_{j}^{2}(t)),
\end{align*}
as indicated by the curly brace in \prettyref{eq:mixed_avg_pair}.

Taken together, we can therefore rewrite the generating functional
\prettyref{eq:Z_pair} averaged over the couplings as
\begin{align}
\bar{Z}[\{\bl^{\alpha},\tbj^{\alpha}\}_{\alpha\in\{1,2\}}] & :=\langle Z[\{\bl^{\alpha},\tbj^{\alpha}\}_{\alpha\in\{1,2\}}](\bJ)\rangle_{\bJ}\label{eq:Zbar_pair_HS}\\
 & =\Pi_{\alpha=1}^{2}\left\{ \int\D Q_{1}^{\alpha}\int\D Q_{2}^{\alpha}\right\} \int\D T_{1}\int\D T_{2}\,\exp\Big(\Omega[\{Q_{1}^{\alpha},Q_{2}^{\alpha}\}_{\alpha\in\{1,2\}},T_{1},T_{2}]\Big)\nonumber \\
\Omega[\{Q_{1}^{\alpha},Q_{2}^{\alpha}\}_{\alpha\in\{1,2\}},T_{1},T_{2}] & :=-\sum_{\alpha=1}^{2}Q_{1}^{\alpha\T}Q_{2}^{\alpha}-T_{1}^{\T}T_{2}+\ln\,Z^{12}[\{Q_{1}^{\alpha},Q_{2}^{\alpha}\}_{\alpha\in\{1,2\}},T_{1},T_{2}]\nonumber \\
Z^{12}[\{Q_{1}^{\alpha},Q_{2}^{\alpha}\}_{\alpha\in\{1,2\}},T_{1},T_{2}] & =\Pi_{\alpha=1}^{2}\Big\{\int\D\bx^{\alpha}\int\D\tbx^{\alpha}\,\exp\Big(S_{0}[\bx^{\alpha},\tbx^{\alpha}]+\bl^{\alpha\T}\bx^{\alpha}+\tbj^{\alpha\T}\tbx^{\alpha}+\tbx^{\alpha\T}Q_{1}^{\alpha}\tbx^{\alpha}+\frac{g^{2}}{2N}\phi(\bx{}^{\alpha})^{\T}Q_{2}^{\alpha}\phi(\bx{}^{\alpha})\Big)\Big\}\nonumber \\
 & \times\exp\left(\tbx^{1\T}\left(T_{1}+D\right)\tbx^{2}+\frac{g^{2}}{N}\phi(\bx^{1})^{\T}T_{2}\phi(\bx^{2})\Big)\right).\nonumber 
\end{align}
We now determine, for vanishing sources, the fields $Q_{1}^{\alpha}$,
$Q_{2}^{\alpha}$, $T_{1}$, $T_{2}$ at which the contribution to
the integral is maximal by requesting $\frac{\delta\Omega}{\delta Q_{1,2}^{\alpha}}=\frac{\delta\Omega}{\delta T_{1,2}}\stackrel{!}{=}0$
for the exponent $\Omega$ of \eqref{eq:Zbar_pair_HS}. Here again
the term $\ln\,Z^{12}$ plays the role of a cumulant generating function
and the fields $Q_{1}^{\alpha},Q_{2}^{\alpha},T_{1},T_{2}$ play the
role of sources, each bringing down the respective factor they multiply.
We denote the expectation value with respect to this functional as
$\langle\circ\rangle_{Q^{\ast},T^{\ast}}$ and obtain the self-consistency
equations
\begin{align}
Q_{1}^{\alpha\ast}(s,t) & =\frac{1}{Z^{12}}\,\frac{\delta Z^{12}}{\delta Q_{2}^{\alpha}(s,t)}=\frac{g^{2}}{2N}\,\sum_{j}\langle\phi(x_{j}^{\alpha})\phi(x_{j}^{\alpha})\rangle_{Q^{\ast},T^{\ast}}\label{eq:saddle_pair}\\
Q_{2}^{\alpha\ast}(s,t) & =0\nonumber \\
T_{1}^{\ast}(s,t) & =\frac{1}{Z^{12}}\,\frac{\delta Z^{12}}{\delta T_{2}(s,t)}=\frac{g^{2}}{N}\,\sum_{j}\langle\phi(x_{j}^{1})\phi(x_{j}^{2})\rangle_{Q^{\ast},T^{\ast}}\nonumber \\
T_{2}^{\ast}(s,t) & =0.\nonumber 
\end{align}
The generating functional at the saddle point is therefore
\begin{align}
\bar{Z}^{\ast}[\{\bl^{\alpha},\tbj^{\alpha}\}_{\alpha\in\{1,2\}}] & =\iint\Pi_{\alpha=1}^{2}\D\bx^{\alpha}\D\tbx^{\alpha}\,\exp\Big(\sum_{\alpha=1}^{2}S_{0}[\bx^{\alpha},\tbx^{\alpha}]+\bl^{\alpha\T}\bx^{\alpha}+\tbj^{\alpha\T}\tbx^{\alpha}+\tbx^{\alpha\T}Q_{1}^{\alpha\ast}\tbx^{\alpha}\Big)\times\nonumber \\
 & \times\exp\left(\tbx^{\alpha\T}\left(T_{1}^{\ast}+D\right)\tbx^{\beta}\right).\label{eq:Z_bar_pair_ast}
\end{align}
We make the following observations: 
\begin{enumerate}
\item The two subsystems $\alpha=1,2$ in the first line of \prettyref{eq:Z_bar_pair_ast}
have the same form as in \eqref{eq:Z_bar_star}. This has been expected,
because the absence of any physical coupling between the two systems
implies that the marginal statistics of the activity in one system
cannot be affected by the mere presence of the second, hence also
their saddle points $Q_{1,2}^{\alpha}$ must be the same as in \eqref{eq:Z_bar_star}.
\item The entire action is symmetric with respect to interchange of any
pair of unit indices. So we have reduced the system of $2N$ units
to a system of $2$ units.
\item If the term in the second line  of \eqref{eq:Z_bar_pair_ast} was
absent, the statistics in the two systems would be independent. Two
sources, however, contribute to the correlations between the systems:
The common Gaussian white noise that gave rise to the term $\propto D$
and the non-white Gaussian noise due to a non-zero value of the auxiliary
field $T_{1}^{\ast}(s,t)$.
\item Only products of pairs of fields appear in \prettyref{eq:Z_bar_pair_ast},
so that the statistics of the $x^{\alpha}$ is Gaussian.
\end{enumerate}
As for the single system, we can express the joint system by a pair
of dynamic equations
\begin{align}
\left(\partial_{t}+1\right)x^{\alpha}(t) & =\eta^{\alpha}(t)\quad\alpha\in\{1,2\}\label{eq:_effective_pair_eq}
\end{align}
together with a set of self-consistency equations for the statistics
of the noises $\eta^{\alpha}$ following from \eqref{eq:saddle_pair}
\begin{align}
\langle\eta^{\alpha}(s)\,\eta^{\beta}(t)\rangle & =D\delta(t-s)+g^{2}\,\langle\phi(x^{\alpha}(s))\phi(x^{\beta}(t))\rangle.\label{eqLeffective_pair_noise}
\end{align}
Obviously, this set of equations \eqref{eq:_effective_pair_eq} and
\eqref{eqLeffective_pair_noise} marginally for each subsystem admits
the same solution solution as determined in \prettyref{sub:particle_motion}.
Moreover, the joint system therefore also possesses the fixed point
$x^{1}(t)\equiv x^{2}(t)$, where the activities in the two subsystems
are identical, i.e. characterized by $c^{12}(t,s)=c^{11}(t,s)=c^{22}(t,s)$
and consequently $d(t)\equiv0\forall t$ \prettyref{eq:mean-squared-distance-1}.

We will now investigate if this fixed point is stable. If it is, this
implies that any perturbation of the system will relax such that the
two subsystems are again perfectly correlated. If it is unstable,
the distance between the two systems may increase, indicating chaotic
dynamics.

We already know that the autocorrelation functions in the subsystems
are stable and each obey the equation of motion \eqref{eq:eq_motion_cxx}.
We could use the formal approach, writing the Gaussian action as a
quadratic form and determine the correlation and response functions
as the inverse, or Green's function, of this bi-linear form. Here,
instead we employ a simpler approach: we multiply the equation \prettyref{eq:_effective_pair_eq}
for $\alpha=1$ and $\alpha=2$ and take the expectation value on
both sides, which leads to
\begin{align*}
\left(\partial_{t}+1\right)\left(\partial_{s}+1\right)\langle x^{\alpha}(t)x^{\beta}(s)\rangle & =\langle\eta^{\alpha}(t)\eta^{\beta}(s)\rangle,
\end{align*}
so we get for $\alpha,\beta\in\{1,2\}$
\begin{align}
\left(\partial_{t}+1\right)\left(\partial_{s}+1\right)c^{\alpha\beta}(t,s) & =D\delta(t-s)+g^{2}F_{\phi}\left(c^{\alpha\beta}(t,s),c^{\alpha\alpha}(t,t),c^{\beta\beta}(s,s)\right)\,,\label{eq:diffeq_cab}
\end{align}
where the function $F_{\phi}$ is defined as the Gaussian expectation
value 
\begin{align*}
F_{\phi}(c^{12},c^{1},c^{2}) & :=\E{\phi(x^{1})\phi(x^{2})}
\end{align*}
for the bi-variate Gaussian
\begin{align*}
\begin{pmatrix}x^{1}\\
x^{2}
\end{pmatrix} & \sim\mathcal{N}_{2}\left(0,\begin{pmatrix}c^{1} & c^{12}\\
c^{12} & c^{2}
\end{pmatrix}\right).
\end{align*}
First, we observe that the equations for the autocorrelation functions
$c^{\alpha\alpha}(t,s)$ decouple and can each be solved separately,
leading to the same equation \eqref{eq:eq_motion_cxx} as before.
As noted earlier, this formal result could have been anticipated,
because the marginal statistics of each subsystem cannot be affected
by the mere presence of the respective other system. Their solutions
\begin{align*}
c^{11}(s,t)= & c^{22}(s,t)=c(t-s)
\end{align*}
then provide the ``background'', i.e., the second and third argument
of the function $F_{\phi}$ on the right-hand side, for the equation
for the crosscorrelation function between the two copies. Hence it
remains to determine the equation of motion for $c^{12}(t,s)$.

We first determine the stationary solution $c^{12}(t,s)=k(t-s)$.
We see immediately that $k(\tau)$ obeys the same equation of motion
as $c(\tau)$, so $k(\tau)=c(\tau)$. The distance \prettyref{eq:mean-squared-distance-1}
therefore vanishes. Let us now study the stability of this solution.
We hence need to expand $c^{12}$ around the stationary solution
\begin{align*}
c^{12}(t,s) & =c(t-s)+\epsilon\,k^{(1)}(t,s)\,,\:\epsilon\ll1\,.
\end{align*}
We develop the right hand side of \prettyref{eq:diffeq_cab} into
a Taylor series using eq. (64) and \prettyref{eq:def_f} 
\begin{align*}
F_{\phi}\left(c^{12}(t,s),c_{0},c_{0}\right) & =f_{\phi}\left(c^{12}(t,s),c_{0}\right)\\
 & =f_{\phi}\left(c(t-s),c_{0}\right)+\epsilon\,f_{\phi^{\prime}}\left(c(t-s),c_{0}\right)\,k^{(1)}(t,s)+O(\epsilon^{2}).
\end{align*}
Inserted into \eqref{eq:diffeq_cab} and using that $c$ solves the
lowest order equation, we get the linear equation of motion for the
first order deflection
\begin{align}
\left(\partial_{t}+1\right)\left(\partial_{s}+1\right)\,k^{(1)}(t,s) & =g^{2}f_{\phi^{\prime}}\left(c(t-s),c_{0}\right)\,k^{(1)}(t,s).\label{eq:variational_equation}
\end{align}
In the next section we will determine the growth rate of $k^{(1)}$
and hence, by \eqref{eq:mean-squared-distance-1}
\begin{align}
d(t) & =\underbrace{c^{11}(t,t)}_{c_{0}}+\underbrace{c^{22}(s,s)}_{c_{0}}\underbrace{-c^{12}(t,t)-c^{21}(t,t)}_{-2c_{0}-\epsilon\,k^{(1)}(t,t)}\nonumber \\
 & =-\epsilon\,k^{(1)}(t,t)\label{eq:relation_distance_c_12}
\end{align}
the growth rate of the distance between the two subsystems. The negative
sign makes sense, since we expect in the chaotic state that $c^{12}(t,s)\stackrel{t,s\to\infty}{=}0$,
so $k^{(1)}$ must be of opposite sign than $c>0$.

\subsection{Schr\"odinger equation for the maximum Lyapunov exponent}

We here want to reformulate the equation for the variation of the
cross-system correlation \prettyref{eq:variational_equation} into
a Schr\"odinger equation, as in the original work \citep[eq. 10]{Sompolinsky88_259}.

First, noting that $C_{\phi^{\prime}\phi^{\prime}}(t,s)=f_{\phi^{\prime}}\left(c(t-s),c_{0}\right)$
is time translation invariant, it is advantageous to introduce the
coordinates $T=t+s$ and $\tau=t-s$ and write the covariance $k^{(1)}(t,s)$
as $k(T,\tau)$ with $k^{(1)}(t,s)=k(t+s,t-s)$. The differential
operator $\left(\partial_{t}+1\right)\left(\partial_{s}+1\right)$
with the chain rule $\partial_{t}\to\partial_{T}+\partial_{\tau}$
and $\partial_{s}\to\partial_{T}-\partial_{\tau}$ in the new coordinates
is $(\partial_{T}+1)^{2}-\partial_{\tau}^{2}$. A separation ansatz
$k(T,\tau)=e^{\frac{1}{2}\kappa T}\,\psi(\tau)$ then yields the eigenvalue
equation
\begin{align*}
(\frac{\kappa}{2}+1)^{2}\psi(\tau)-\partial_{\tau}^{2}\psi(\tau) & =g^{2}f_{\phi^{\prime}}\left(c(\tau),c_{0}\right)\psi(\tau)
\end{align*}
for the growth rates $\kappa$ of $d(t)=-k^{(1)}(t,t)=-k(2t,0)$.
We can express the right hand side by the second derivative of the
potential \prettyref{eq:def_potential_V} $V(c(\tau);c_{0})$ so that
with 
\begin{eqnarray}
V^{\prime\prime}(c(\tau);c_{0}) & = & -1+g^{2}f_{\phi^{\prime}}\left(c(\tau),c_{0}\right)\label{eq:effective_potential}
\end{eqnarray}
we get the time-independent Schrödinger equation
\begin{eqnarray}
\left(-\partial_{\tau}^{2}-V^{\prime\prime}(c(\tau);c_{0})\right)\psi(\tau) & = & \underbrace{\left(1-\left(\frac{\kappa}{2}+1\right)^{2}\right)}_{=:E}\psi(\tau).\label{eq:Schroedinger}
\end{eqnarray}
The eigenvalues (``energies'') $E_{n}$ determine the exponential
growth rates $\kappa_{n}$ the solutions $k(2t,0)=e^{\kappa_{n}t}\,\psi_{n}(0)$
at $\tau=0$ with 
\begin{eqnarray}
\kappa_{n}^{\pm} & = & 2\left(-1\pm\sqrt{1-E_{n}}\right).\label{eq:roots_lambda}
\end{eqnarray}
We can therefore determine the growth rate of the mean-square distance
of the two subsystems in \prettyref{sub:pair_of_systems} by \prettyref{eq:relation_distance_c_12}.
The fastest growing mode of the distance is hence given by the ground
state energy $E_{0}$ and the plus in \prettyref{eq:roots_lambda}.
The deflection between the two subsystems therefore growth with the
rate
\begin{eqnarray}
\Lambda_{\mathrm{max}} & = & \frac{1}{2}\kappa_{0}^{+}\label{eq:Lambda_max}\\
 & = & -1+\sqrt{1-E_{0}},\nonumber 
\end{eqnarray}
where the factor $1/2$ in the first line is due to $d$ being the
squared distance, hence the length $\sqrt{d}$ growth with half the
exponent as $d$.

\begin{figure}
\begin{centering}
\includegraphics[scale=2]{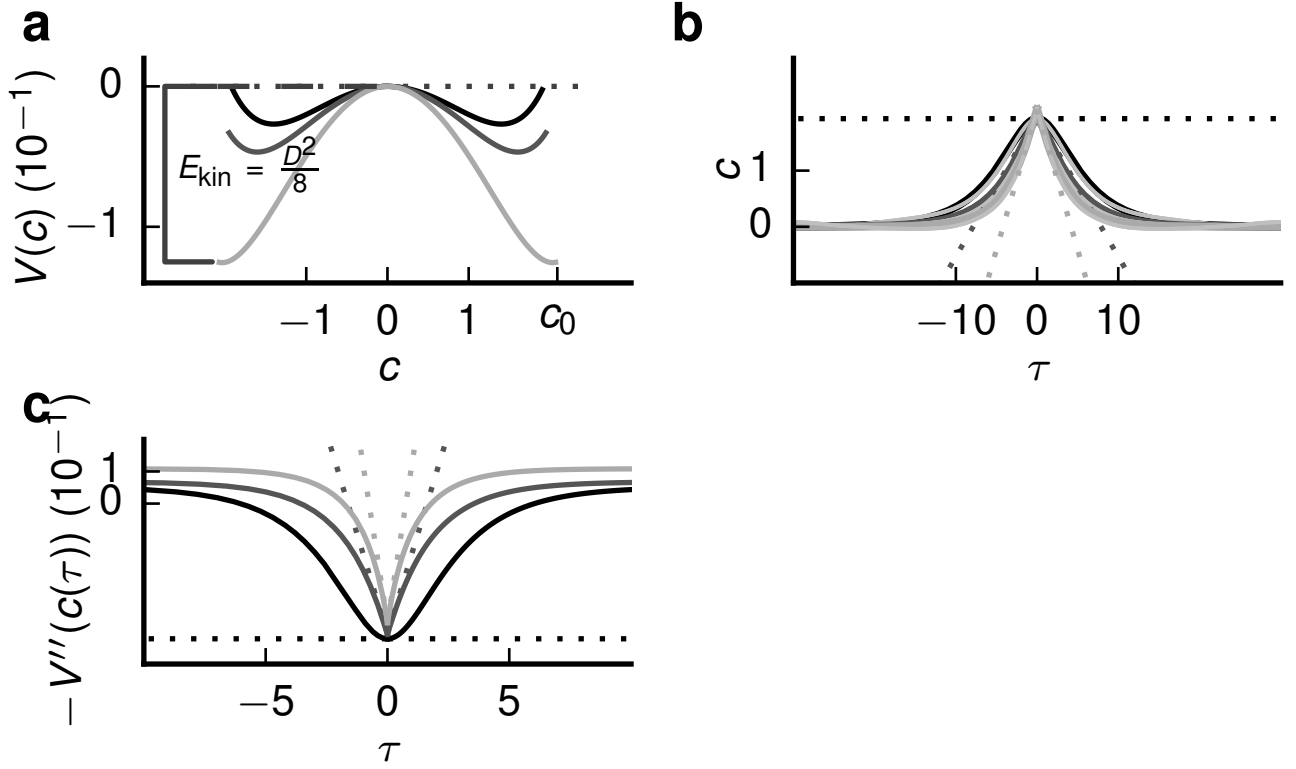}
\par\end{centering}
\caption{\textbf{Dependence of the self-consistent solution on the noise level
$D$.} \textbf{a} Potential that determines the self-consistent solution
of the autocorrelation function \eqref{eq:def_potential_V}. Noise
amplitude $D>0$ corresponds to an initial kinetic energy $E_{\mathrm{kin}}=\frac{D^{2}}{8}$.
The initial value $c_{0}$ is determined by the condition $V(c_{0};c_{0})+E_{\mathrm{kin}}=0$,
so that the ``particle'' starting at $c(0)=c_{0}$ has just enough
energy to reach the peak of the potential at $c(\tau\to\infty)=0$.
In the noiseless case, the potential at the initial position $c(0)=c_{0}$
must be equal to the potential for $\tau\to\infty$, i.e. $V(c_{0};c_{0})=V(0)=0$,
indicated by horizontal dashed line and the corresponding potential
(black). \textbf{b} Resulting self-consistent autocorrelation functions
given by \prettyref{eq:acf_system}. The kink at zero time lag $\dot{c}(0-)-\dot{c}(0+)=\frac{D}{2}$
is indicated by the tangential dotted lines. In the noiseless case
the slope vanishes (horizontal dotted line). Simulation results shown
as light gray underlying curves.\textbf{ c} Quantum mechanical potential
appearing in the Schr\"odinger equation \prettyref{eq:Schroedinger}
with dotted tangential lines at $\tau=\pm0$. Horizontal dotted line
indicates the vanishing slope in the noiseless case. Other parameters
as in \prettyref{fig:effective_potential}.\label{fig:noise_dependence}}
\end{figure}
Energy conservation \prettyref{eq:initial_c0} determines $c_{0}$
also in the case of non-zero noise $D\neq0$, as shown in \prettyref{fig:noise_dependence}a.
The autocovariance function obtained from the solution of \prettyref{eq:acf_system}
agrees well to the direct simulation \prettyref{fig:noise_dependence}b.
The quantum potential appearing in \prettyref{eq:Schroedinger} is
shown in \prettyref{fig:noise_dependence}c.

\subsection{Condition for transition to chaos}

We can construct an eigensolution of \prettyref{eq:Schroedinger}
from \prettyref{eq:eq_motion_cxx}. First we note that for $D\neq0$,
$c$ has a kink at $\tau=0$. This can be seen by integrating \prettyref{eq:eq_motion_cxx}
from $-\epsilon$ to $\epsilon$, which yields
\begin{eqnarray*}
\lim_{\epsilon\to0}\int_{-\epsilon}^{\epsilon}\partial_{\tau}^{2}cd\tau & = & \dot{c}(0+)-\dot{c}(0-)\\
 & = & D.
\end{eqnarray*}
Since $c(\tau)=c(-\tau)$ is an even function it follows that $\dot{c}(0+)=-\dot{c}(0-)=-\frac{D}{2}$.
For $\tau\neq0$ we can differentiate \prettyref{eq:eq_motion_cxx}
with respect to time $\tau$ to obtain 
\begin{eqnarray*}
\partial_{\tau}\partial_{\tau}^{2}\,c(\tau) & = & \partial_{\tau}^{2}\,\dot{c}(\tau)\\
=-\partial_{\tau}V^{\prime}(c(\tau)) & = & -V^{\prime\prime}(c(\tau))\,\dot{c}(\tau).
\end{eqnarray*}
Comparing the right hand side expressions shows that $\left(\partial_{\tau}^{2}+V^{\prime\prime}(c(\tau))\right)\dot{c}(\tau)=0$,
so $\dot{c}$ is an eigensolution for eigenvalue $E_{n}=0$ of \prettyref{eq:Schroedinger}.

Let us first study the case of vanishing noise $D=0$ as in \citep{Sompolinsky88_259}.
The solution then $\dot{c}$ exists for all $\tau$. Since $c$ is
a symmetric function, $\Psi_{0}=\dot{c}$ has single node. The single
node of this solution implies there must be a state with zero nodes
that has even lower energy, i.e. $E_{0}<0$ . This, in turn, indicates
a positive Lyapunov exponent $\Lambda_{\mathrm{max}}$ according to
\prettyref{eq:Lambda_max}. This is the original argument in \citep{Sompolinsky88_259},
showing that at $g=1$ a transition from a silent to a chaotic state
takes place.

Our aim is to find the parameter values for which the transition to
the chaotic state takes place in the presence of noise. We know that
the transition takes place if the eigenvalue of the ground state of
the Schr\"odinger equation is zero. We can therefore explicitly try
to find a solution of \prettyref{eq:Schroedinger} for eigenenergy
$E_{n}=0$, i.e. we seek the homogeneous solution that satisfies all
boundary conditions, i.e. continuity of the solution as well as its
first and second derivative. We already know that $\dot{c}(\tau)$
is one homogeneous solution of \prettyref{eq:Schroedinger} for positive
and for negative $\tau$. For $D\neq0$, we can construct a continuous
solution from the two branches by defining

\begin{eqnarray}
y_{1}(\tau) & = & \begin{cases}
\dot{c}(\tau) & \tau\ge0\\
-\dot{c}(\tau) & \tau<0
\end{cases},\label{eq:def_y1}
\end{eqnarray}
which is symmetric, consistent with the search for the ground state.
In general $y_{1}$ does not solve the Schr\"odinger equation, because
the derivative at $\tau=0$ is not necessarily continuous, since by
\eqref{eq:diffeq_auto} $\partial_{\tau}y_{1}(0+)-\partial_{\tau}y_{1}(0-)=\ddot{c}(0+)+\ddot{c}(0-)=2(c_{0}-g^{2}f_{\phi}(c_{0};c_{0}))$.
Therefore $y_{1}$ is only an admissible solution, if the right hand
side vanishes. The criterion for the transition to the chaotic state
is hence

\begin{align}
0=\partial_{\tau}^{2}c(0\pm) & =c_{0}-g^{2}f_{\phi}\left(c_{0},c_{0}\right)\label{eq:ch_trans_noise}\\
 & =-V^{\prime}(c_{0};c_{0}).\nonumber 
\end{align}
The latter condition therefore shows that the curvature of the autocorrelation
function vanishes at the transition. In the picture of the motion
of the particle in the potential the vanishing acceleration at $\tau=0$
amounts to a potential with a flat tangent at $c_{0}$.

The criterion for the transition can be understood intuitively. The
additive noise increases the peak of the autocorrelation at $\tau=0$.
In the large noise limit, the autocorrelation decays as $e^{-|\tau|}$,
so the curvature is positive. The decay of the autocorrelation is
a consequence of the uncorrelated external input. In contrast, in
the noiseless case, the autocorrelation has a flat tangent at $\tau=0$,
so the curvature is negative. The only reason for its decay is the
decorrelation due to the chaotic dynamics. The transition between
these two forces of decorrelation hence takes place at the point at
which the curvature changes sign, from dominance of the external sources
to dominance of the intrinsically generated fluctuations. For a more
detailed discussion please see \citep{Goedeke16_arxiv}.

\section{Appendix}

\subsection{Price's theorem\label{sub:prices_theorem}}

We here provide a derivation of Price's theorem \citep{PapoulisProb},
which, for the Gaussian integral \prettyref{eq:def_f} takes the
form
\begin{align}
\frac{\partial}{\partial c}f_{u}(c,c_{0}) & =f_{u^{\prime}}(c,c_{0}).\label{eq:prices}
\end{align}
We here provide a proof using the Fourier representation of $u(x)=\frac{1}{2\pi}\int\,U(\omega)\,e^{i\omega x}d\omega$.
Alternatively, integration by parts can be used to obtain the same
result by a slightly longer calculation. We write the integral as
\begin{align*}
f_{u}(c,c_{0}) & =\iint\iint\,U(\omega)\,e^{i\omega\left(\frac{1}{\sqrt{c_{0}}}\sqrt{c_{0}^{2}-c^{2}}z_{1}+\tfrac{c}{\sqrt{c_{0}}}z_{2}\right)}U(\omega^{\prime})\,e^{i\omega^{\prime}\left(\sqrt{c_{0}}z_{2}\right)}\,Dz_{1}Dz_{2}\;d\omega d\omega^{\prime}\\
 & =\iint\,U(\omega)\,e^{\frac{1}{2}\frac{1}{c_{0}}\left(c_{0}^{2}-c^{2}\right)\omega^{2}}U(\omega^{\prime})\,e^{\frac{1}{2}\frac{1}{c_{0}}\left(c\omega+c_{0}\omega^{\prime}\right)^{2}}\;d\omega d\omega^{\prime},
\end{align*}
where we used the characteristic function $e^{\frac{1}{2}\omega^{2}}$
of the unit variance Gaussian contained in the measures $Dz$. The
derivative by $c$ with the product rule yields
\begin{align*}
\frac{\partial}{\partial c}f_{u}(c,c_{0}) & =\iint\,\left(-\frac{c}{c_{0}}\omega^{2}+\frac{c\omega+c_{0}\omega^{\prime}}{c_{0}}\omega\right)U(\omega)\,\,e^{\frac{1}{2}\frac{1}{c_{0}}\left(c_{0}^{2}-c^{2}\right)\omega^{2}}U(\omega^{\prime})\,e^{\frac{1}{2}\frac{1}{c_{0}}\left(c\omega+c_{0}\omega^{\prime}\right)^{2}}\;d\omega d\omega^{\prime}\\
 & =\iint\,\omega\omega^{\prime}\,U(\omega)\,\,e^{\frac{1}{2}\frac{1}{c_{0}}\left(c_{0}^{2}-c^{2}\right)\omega^{2}}U(\omega^{\prime})\,e^{\frac{1}{2}\frac{1}{c_{0}}\left(c\omega+c_{0}\omega^{\prime}\right)^{2}}\;d\omega d\omega^{\prime}\\
 & =f_{u^{\prime}}(c,c_{0}),
\end{align*}
where we used $u^{\prime}(x)=\frac{1}{2\pi}\int\,i\omega\,U(\omega)\,e^{i\omega x}d\omega$
in the last step, proving the assertion.

\begin{acknowledgments}
The authors are thankful for helpful discussions with Andrea Crisanti.
This work was partly supported by the Helmholtz association: VH-NG-1028
and SMHB; EU Grant 604102 (HBP), Juelich Aachen Research Alliance
(JARA). 
\end{acknowledgments}

\bibliographystyle{apsrev_brain}
\bibliography{brain,math,computer}

\end{document}